\documentclass[10pt, letter, twocolumn]{IEEEtran}

\usepackage{color}
\usepackage{epsf}
\usepackage{times}
\usepackage{epsfig}
\usepackage{graphicx}
\usepackage{algorithm}
\usepackage{algorithmic}
\usepackage{amsmath}
\usepackage{amssymb}
\usepackage{amsxtra}
\usepackage{mathtools}
\usepackage{amsfonts}
\usepackage{cite}
\usepackage{cleveref}
\DeclarePairedDelimiter{\ceil}{\lceil}{\rceil}
\usepackage{bbm}
\usepackage{caption}
\usepackage{subcaption}
\usepackage{multirow}
\usepackage{array}
\newcolumntype{P}[1]{>{\centering\arraybackslash}p{#1}}

\begin{document}
\title{
Green Virtualization for Multiple Collaborative Cellular Operators
}
\author{\IEEEauthorblockN{\vspace{-0.0cm}Muhammad Junaid Farooq, \textit{Student Member, IEEE}, \large Hakim Ghazzai, \textit{Member, IEEE}, Elias Yaacoub, \textit{Senior Member, IEEE}, Abdullah Kadri, \textit{Senior Member, IEEE}, and Mohamed-Slim Alouini, \textit{Fellow, IEEE}\vspace{-0.0cm}}\\
\vspace{-0.2cm}
\thanks{\hrule \vspace{0.2cm}
This work was made possible by grant NPRP \# 6-001-2-001 from the Qatar National Research Fund (a member of The Qatar Foundation). The statements made herein are solely the responsibility of the authors.

Muhammad Junaid Farooq, Hakim Ghazzai, and Abdullah Kadri are with Qatar Mobility Innovations Center (QMIC), Qatar University, Doha, Qatar. E-mails: \{junaidf, hakimg, abdullahk\}@qmic.com.

Elias Yaacoub is with Faculty of Computer Studies, Arab Open University (AOU), Beirut, Lebanon. E-mail: eliasy@ieee.org.

Mohamed-Slim Alouini is with King Abdullah University of Science and Technology (KAUST), Thuwal, Makkah Province, Saudi Arabia. E-mail: slim.alouini@kaust.edu.sa.

This work was done while Muhammad Junaid Farooq was at QMIC. He is now with the Tandon School of Engineering, New York University, Brooklyn, NY, United States. E-mail: mjf514@nyu.edu.
}}

\maketitle
\thispagestyle{empty}

\begin{abstract}
\boldmath
This paper proposes and investigates a green virtualization framework for infrastructure sharing among multiple cellular operators whose networks are powered by a combination of conventional and renewable sources of energy. Under the proposed framework, the virtual network formed by unifying radio access infrastructures of all operators is optimized for minimum energy consumption by deactivating base stations (BSs) with low traffic loads. The users initially associated to those BSs are off-loaded to neighboring active ones. A fairness criterion for collaboration based on roaming prices is introduced to cover the additional energy costs incurred by host operators. The framework also ensures that any collaborating operator is not negatively affected by its participation in the proposed virtualization. A multi-objective linear programming problem is formulated to achieve energy and cost efficiency of the networks' operation by identifying the set of inter-operator roaming prices. For the case when collaboration among all operators is infeasible due to profitability, capacity, or power constraints, an iterative algorithm is proposed to determine the groups of operators that can viably collaborate. Results show significant energy savings using the proposed virtualization as compared to the standalone case. Moreover, collaborative operators exploiting locally generated renewable energy are rewarded more than traditional ones.
\end{abstract}

\begin{IEEEkeywords}
\noindent Base station sleeping, energy efficiency, green networking, mobile operator collaboration, roaming price.
\end{IEEEkeywords}
\section{Introduction}
The advent of third generation (3G) and fourth generation (4G) mobile broadband technologies has led to a multi-fold increase in the number of mobile devices and this trend is expected to tremendously accentuate with the emergence of fifth generation (5G) mobile networks. Currently, the number of mobile devices exceeds the number of people on earth (i.e., approximately 7.6 billion) and is expected to increase to around 11.5 billion in 2019~\cite{cisco_report}. The increasing use of data and cloud-based services in the future will require significant capacity enhancement efforts by cellular operators including the deployment of additional base stations (BSs) to serve more users. This will enormously inflate energy consumption of the cellular networks. Consequently, the fossil fuel consumption will increase since they are expected to continue supplying about 80\% of the world energy through 2040~\cite{weo}. This will lead to the increase in emission of pollutant gases into the atmosphere mainly the carbon dioxide (CO$_2$)~\cite{ericsson_report,7570259}. Furthermore, such high energy consumption will force cellular operators to pay huge energy bills that form a major portion of their operational expenditures (OPEX). Currently, the worldwide annual OPEX of cellular operators for electricity is more than $\$10$ billion \cite{energy_efficiency_need}. Therefore, there is a pressing need to reduce energy consumption of cellular networks by efficient utilization of communication infrastructure for the sake of their own profitability as well as the environment.

Cellular networks are generally controlled by multiple mobile operators within a well-defined geographical area. Each operator deploys its radio access network (RAN) after careful consumer demand assessment and planning. BSs belonging to different operators are usually located independently of each other although they may sometimes be co-located, e.g., on top of the same building. Nevertheless, there is always an overlap of coverage regions of different operators. This leads to redundancy in energy utilization of the communication network as almost half of the power consumption of BSs, which are the most power hungry components~\cite{manifesto}, is un-utilized in low traffic periods. Therefore, existing studies aiming to reduce energy consumption of cellular networks~\cite{green_survey,green_survey_2,cell_zooming} focus primarily on the RAN, i.e., BSs. Motivated by this geographical co-existence, the concept of infrastructure sharing between multiple operators, also known as wireless network virtualization~\cite{network_virtualization}, is introduced for efficient energy utilization~\cite{infrastructure_sharing_report,infrastructure_sharing}. Initially, this concept was proposed for efficient spectrum utilization and OPEX reduction. In the context of green networking, virtualization is used to reduce energy consumption and consequently the environmental impact of cellular networks, thus referred to as \emph{green virtualization}. According to a European study~\cite{network_sharing}, energy consumption of mobile networks can be reduced by up to 60$\%$ if communication infrastructure is shared by mobile operators. Several types of infrastructure sharing exist such as site sharing, mast sharing, RAN sharing, network roaming, and core network sharing~\cite{infrastructure_sharing_report}. In this paper, we focus on the network roaming type sharing also referred to as roaming-based infrastructure sharing~\cite{infrastructure_sharing}.

Under the roaming-based infrastructure sharing scheme, mobile operators pool their RAN resources to form a virtual network. Each mobile operator relies on the coverage of other operators' networks to serve its users in regions that are out of its own coverage. This means that the users of all operators can connect to any BS belonging to any operator subject to suitable roaming agreements. In other words, each operator now virtually owns the combined set of BSs of all operators and can use their resources to serve its customers. This infrastructure sharing is henceforth referred to as \emph{collaboration} since operators cognitively join hands to serve the collective users for mutual benefit~\cite{7294664}. \textcolor{black}{Resource sharing and coalition formation for common interests is also used in other settings such as in the case of multiple cloud providers~\cite{greedy_ref_3} and networks using device-to-device communications~\cite{greedY_ref_2}}. Due to the flexibility of serving users, an operator may switch off its lightly loaded BS while intelligently offloading the associated users to BSs of other operators. Optimally, only a portion of each operator's network will be active at one time and together, they serve the entire geographical region while minimizing the energy consumption. This requires cognitive decision-making for forming collaborative groups and turning off redundant BSs based on the situational awareness of the network and the users.
\subsection{Literature Review}
\label{sec2}
The first step towards green cellular networks is the energy efficient deployment of BSs. During this procedure, a complete demand assessment is made and the coverage overlap is minimized to reduce the interference and subsequently, the transmit power needed to meet QoS requirements. In this regard, efforts have been made for effective cell planning~\cite{7056465} and BS deployment~\cite{deployment1}. Nevertheless, the BS deployment is done according to estimates of peak demand levels. However, the user load of BSs may change during different times of the day, which may lead to under-utilization of power and communication resources.

To further improve energy efficiency, several techniques have been used in literature to cater for load variations across the deployed cellular networks. One of these techniques is cell range adaptation~\cite{range_adaptation}, also known as cell zooming~\cite{cell_zooming}, in which the transmit power of the BSs is adjusted according to the number of connected users. Therefore, the lightly loaded BSs transmit with lower power and hence, have a smaller range as compared to BSs with higher number of connected users.

Another important technique used for energy saving in modern cellular networks is the BS sleeping strategy. This is similar to cell zooming except that the lightly loaded BSs are completely turned off and the connected users are off-loaded to nearby active BSs. The BS sleeping strategy is extensively studied in literature for single operator cellular systems~\cite{greedy_Bousia,6489498,BS_sleeping_12,BS_sleeping_11}. A framework for switching off BSs in cellular networks based on the distances between BSs and users is proposed in ~\cite{greedy_Bousia}. In~\cite{6489498}, a dynamic BS switching algorithm is proposed in which BSs are turned off one at a time to have minimum impact on the rest of the network. The results show 50 to 80$\%$ energy saving based on real traffic profile from a metropolitan area. In~\cite{BS_sleeping_12}, the sleeping control of a single BS in a cellular network is optimized. Different BS sleeping modes are discussed and the mode that achieves the Pareto optimal tradeoff between total power consumption and average delay is selected. In~\cite{BS_sleeping_11}, a distributed cooperative framework for BS sleeping in cellular networks is proposed. In this framework, the neighboring BSs cooperate to optimize the BS sleeping strategy based on the traffic load. The energy minimization problem in this cooperative setting is formulated as a constrained graphical game and solved using a game theoretic solution.

It is pertinent to mention here that several other techniques for energy efficiency are also widely used in practice such as the deployment of low powered small cell BSs in areas of high demand instead of the macro BS~\cite{small_cell_green}. However, the energy saving potential of optimizing the consumption of macrocell BSs is much higher than that of using small cell BSs. Therefore, only macro BSs are considered in this paper. The current research interest is moving towards achieving energy efficiency in multi-operator cellular networks. In such networks, there is an additional degree of freedom for achieving energy efficiency apart from the previously mentioned techniques. This is related to the operators' virtualization that allows turning off more BSs as compared to the case when each operator is in a standalone setting. Several efforts have been made in literature to study the collaboration between multiple network operators~\cite{coop2,coop3,coop4}. In~\cite{coop2}, the energy saving potential of the collaboration among multiple operators serving the same area is evaluated. The traffic load of the network is successively reduced from the peak level and the BS switching ON/OFF behavior of different network deployments is studied. It is shown that significant energy savings can be achieved if collaboration is enabled among operators. In~\cite{coop3}, the economic benefits of multi-operator collaboration have been studied. Significant reduction in OPEX of collaborating operators has been shown. Practical effort towards optimizing the BS sleeping process in multi-operator environments are made in~\cite{coop4,7417231} where a game theoretic approach is proposed in~\cite{coop4} to select the active BSs combination that minimizes the OPEX of operators while in,~\cite{7417231}, a cooperative heuristic approach is adopted to turn off redundant BSs. The need to introduce fairness and stability in the collaboration framework is discussed in~\cite{6347622} and a game theoretic model is also provided. Most of the discussed frameworks are based on instantaneous network statistics which results in an optimized solution for a single time instant. However, the obtained solution is highly sensitive to variations in network parameters such as channel realizations, traffic, etc. Hence, the optimized instantaneous solution might not be reliable and useful as a guideline for long term planning and/or mutual agreements between the operators. In contrast, this study focuses on using average network statistics which can provide more generalized results.

Finally, the use of renewable energy (RE) for cellular networks has also been studied in detail in the literature~\cite{renewable,BS_sleeping_renewable}. A survey on the RE usage in cellular networks is provided in~\cite{renewable} that presents the different architectures for incorporating green energy in the power supply of cellular equipment. The most widely accepted model is to generate RE at cellular BS sites. As an example,~\cite{BS_sleeping_renewable} utilizes RE generated at BS sites into the BS sleeping framework in order to achieve both energy and cost efficiency. Another work, presented in~\cite{6778102}, proposed the optimization of the energy procurement, particularly the green energy, of mobile operators in smart grid environment.

\subsection{Contributions}
\textcolor{black}{Most of the aforementioned studies in multi-operator collaboration are either limited to two mobile operators~\cite{coop4}, which is a simplified version of the general multi-operator case, or do not consider the collaboration cost~\cite{network_sharing,coop2}. The ones that consider the multi-operator case assume that BSs of the operators are collocated~\cite{7417231,7124517}, which ignores the spatial distribution of the users/BSs and its effects on the infrastructure sharing. The collaboration between more than two operators is non-trivial since each operator will have to manage separate agreements with the remaining operators based on the number of their roamed users.} Indeed, collaboration among mobile operators provides more flexibility in achieving energy efficiency as compared to the traditional scenario. However, uncontrolled collaboration would be unfair for one or multiple operators. For instance, one operator might turn off all its BSs so that all its users are roamed to the infrastructure of the other operators. As a result, the serving operators might suffer from high energy consumption while the first operator enjoys its profit increase. Although this solution may maximize the overall objective of operators, the individual objective distribution is unfair. Therefore, it is important to introduce some fairness criteria during the collaboration process that will influence the collaboration decision of operators. \textcolor{black}{It can take different forms such as the collaboration under equal charge allocation (ECA) where the total cost is equally shared among operators~\cite{6347622}. Another fairness criterion could be the proportional share of the collaboration cost with respect to the number of users, also known as the cost gap allocation (CGA) in \cite{6347622}}. In this case, only the cost due to collaboration is considered and shared among operators. Note that these fairness criteria force operators to be members of the collaboration group and share the cost with other competitive operators.

\textcolor{black}{In this study, we propose a generalized virtualization framework for existing networks of standalone mobile operators to collaborate and achieve energy saving. The developed framework aims to select an effective active BS combination for the collaborative networks to serve the existing users. A fairness criterion based on roaming prices defined by each operator who is willing to serve users of competitive operators is adopted. The proposed method will also allow operators to decide whether to enter into collaboration or not after checking if their profitability will not be affected due to collaboration.}
\textcolor{black}{Furthermore, we incorporate the availability of locally generated RE, e.g., solar, wind, etc., at BSs' sites in the optimization framework. Unlike other studies where mobile operators collaborate in a competitive manner in order to minimize their own operational profits even though they may turn off some BSs to reduce the energy cost~\cite{7124517}, the proposed framework advocates a fair green collaborative virtualization solution where the joint objective of mobile operators is to minimize the fossil fuel consumption.} The main contributions of this paper can be summarized as follows:
\begin{itemize}
  \item \textcolor{black}{A novel optimization problem aiming at minimizing the energy consumption of operators while fulfilling their BS power budget and capacity constraints is formulated. Profitability constraints are also introduced to ensure that each operator is willing to collaborate only if its profit after collaboration is equal to or higher than its profit before collaboration.}
\item \textcolor{black}{The concept of user off-loading by operators from their BSs to those of other operators in exchange for a charge (i.e., roaming price) for each roamed customer is employed. The roaming decision is made by solving a multi-objective linear programing problem (MOLPP) and is taken at the operator level such that users experience seamless connectivity.}
  \item An iterative algorithm is proposed to assess the possibility of collaboration between operators and to split them into multiple groups of collaborating operators. This is useful when joint collaboration among all operators is not possible. The effect of RE production on the collaboration decision and the subsequent roaming prices is also investigated in this work.
  \item Average network statistics such as the average number of users in each cell, the average transmit power of the BSs, and the average cost of fossil fuel, etc., are considered in the optimization framework. Therefore, the resulting average roaming price decisions can be used as a reliable guideline for operators while making collaboration agreements.
\end{itemize}

The rest of the paper is organized as follows: Section~\ref{sec3} presents the system model. The problem formulation for the non-collaborative and collaborative scenarios are described in Section~\ref{sec4}. The proposed green multi-operator collaborative solution in conjunction with the BS sleeping strategy algorithm is detailed in Section~\ref{green_solution}. Simulation results and future challenges in multi-operator collaboration are given in Section~\ref{sec7}. Finally, Section~\ref{sec9} concludes the paper.
\section{System Model}
\label{sec3}
We consider a geographical area in $\mathbb{R}^{2}$ denoted by $\mathcal{A}$ and served by $N_{\text{op}}$ mobile operators. Each operator $l=1,\cdots, N_{\text{op}}$ deploys a cellular network that satisfies the QoS of its customers, i.e., traffic demands, and covers the total area. $N_{\text{BS}}^{(l)}$ denotes the number of BSs that are deployed by operator~$l$~in that area according to a well planned strategy.

\subsection{User Distribution}
The network's users are connected to the nearest available BS, therefore, the coverage region of all BSs forms a Voronoi tessellation~\cite{voronoi} where each cell depicts the region in space in which all users are associated with the corresponding BS. If the operators do not collaborate, the coverage regions and the user association will be independent of each other. However, if the operators choose to collaborate, the coverage regions of the BSs will be different and users of one operator will be able to connect to BSs of the other operators. Hence, the concept of roaming comes into effect. \textcolor{black}{It should be noted here that the Voronoi cells are formed in accordance to the user-BS association method which is based on minimum distances. As this study is based on the average statistics of the networks, the users are assigned to the BSs according to their path loss levels in order to minimize the BSs' radiated power levels. This does not induce that the cell overlaps among the BSs that belong to the same and/or other operators are ignored. On the contrary, cell overlap is an important factor for infrastructure sharing so that users can be off-loaded to neighbor BSs.}

\textcolor{black}{Let $N_{U}^{(l)}$ be the total number of users connected to network $l$ benefiting from one of the different services provided by each operator, denoted by $\Sigma_l, l=1,\cdots,N_{\text{op}}$. We denote by $N_{U}^{(l,\sigma)}$ the number of users of network $l$ using the service $\sigma$ offered by operator $l$ where $\sigma=1,\cdots, \Sigma_l$ such that $\sum_{\sigma=1}^{\Sigma_l}N_{U}^{(l,\sigma)}=N_{U}^{(l)}$. The users of network $l$ using service $\sigma$ are placed according to a given joint probability density function (pdf) in the total region $\mathcal C$ denoted by $f_{l,\sigma}(x,y)$. For instance, the density could follow a uniform distribution with a given user density per km$^2$ or a normal distribution corresponding to concentrated users in a hotspot area and then, the density is reduced as we move away from the center, etc.
The proportion of users using service $\sigma$ in a sub-region $C$, i.e., $C \subseteq \mathcal C$, is computed as $\iint_C f_{l,\sigma}(x,y) dx\,dy$. Thus, the total number of users of network $l$ using service $\sigma$ in $C$ is denoted by $N_{C}^{(l,\sigma)}$ and is given by:
\begin{equation}
\label{Usereq}
N_{C}^{(l,\sigma)}=\ceil{N_{U}^{(l,\sigma)}\iint_C f_{l,\sigma}(x,y) dx\,dy},
\end{equation}
where $\ceil{.}$ denotes the ceiling function.}

\subsection{Energy Consumption Model of Base Stations}
\textcolor{black}{We consider that each BS is equipped with a single omni-directional antenna and serves a cell $A_{j}^{(l)}$, where $A_{j}^{(l)}$ denotes the Voronoi cell corresponding to the $j^{th}$ BS of operator $l$. If a BS $j$ of operator $l$ is completely switched off then, we assume that its average power consumption $P_{j}^{(l)}=0$. However, the average consumed power of a switched on BS $j$ belonging to operator $l$ is computed as follows~\cite{TVTjournal1}:
\begin{equation}
P_j^{(l)}=a P_{\text{tx},j}^{(l)}+b,
\label{equationpowermodel}
\end{equation}
where the coefficient $a$ corresponds to the power consumption that scales with the radiated power due to amplifier and feeder losses and the term $b$ models an offset of site power that is consumed independently of the average transmit power and is due to signal processing, battery backup, and cooling, etc. In (\ref{equationpowermodel}), $P_{\text{tx},j}^{(l)}$ denotes the average radiated power of the $j^{th}$ BS that belongs to operator $l$ and can be expressed as follows:
\begin{equation}
\label{transmitPowereq}
P_{\text{tx},j}^{(l)}= \sum_{\sigma=1}^{\Sigma_l}N_{A_{j}^{(l)}}^{(l,\sigma)}\frac{P_{\text{min}}^{\sigma}}{K} \mathbb{E}\left[\left(r_{lj}^{\sigma}\right)^{\eta}\right],
\end{equation}
where $P_{\min}^{\sigma}$ is the minimum received power for effective signal detection and represents the QoS metric of service $\sigma$. In other words, the operator has to ensure that each user using service $\sigma$ receives a minimum power of $P_{\min}^{\sigma}$ in order to meet the rate requirement. In~\eqref{transmitPowereq}, $K$ is the path loss constant, $\eta$ is the path loss exponent, and $r_{lj}^\sigma$ is a random variable denoting the distance between the $j^{th}$ BS and the users of operator $l$ using service $\sigma$ expressed as follows:
\begin{equation}
r_{lj}^\sigma = \sqrt{(x^{\sigma} - x_{lj})^{2} + (y^{\sigma} - y_{lj})^{2}},
\end{equation}
where $(x^\sigma,y^\sigma)$ are the location coordinates of a random user using service $\sigma$ following the distribution $f_{l,\sigma}(x,y)$ and $(x_{lj},y_{lj})$ are the coordinates of the $j^{th}$ BS of operator $l$. The distance function $\left(r_{lj}^{\sigma}\right)^{\eta}$ is averaged over the Voronoi cell $A_{j}^{(l)}$ as follows:
\small
\begin{equation}\label{average_distance}
\mathbb{E}[\left(r_{lj}^{\sigma}\right)^{\eta}] = \iint_{A_{j}^{(l)}} \left( (x^{\sigma} - x_{lj})^{2} + (y^{\sigma} - y_{lj})^{2} \right)^{\eta/2}f_{l,\sigma}(x,y) dx \ dy.
\end{equation}}
\normalsize

To power its BSs, an operator can either procure energy from a traditional electricity provider or use RE generators, e.g., solar panels or wind turbines installed on BS sites. The amount of energy procured from the electricity grid and the amount of RE consumed by BS $j$ of operator $l$ are denoted by $q_{lj}$ and $g_{lj}$, respectively.

The amount of green energy generated locally is varying from one BS to another depending on environmental and/or technical reasons. For instance, the solar rating depends essentially on the size of photovoltaic panels and whether they experience any shading during the~day.
It should be noted that the locally generated RE is free of charge unlike the energy procured from external retailer that is evaluated by the cost of one unit of energy for operator $l$, denoted by $\pi_{l}$. The amount of grid electricity procured by BS $j$ of operator $l$ is given as follows:
\begin{equation}
\label{fossilenergy}
q_{lj}=\max(P_j^{(l)}\Delta t-g_{lj}, 0), \forall j=1,\cdots, N_{\text{BS}}^{(l)}, \forall l=1,\cdots, N_{\text{op}},
\end{equation}
where $\Delta t$ is the BS operation time \textcolor{black}{during the collaboration period}.

In order to differentiate between the variables related to the non-collaborative scenario and those of the collaborative scenario, we employ the following notations in the sequel: $x^{(u)}$ and $x^{(c)}$, where $u$ and $c$ stand for non-collaborative and collaborative, respectively.

\section{Optimization Problems} \label{sec4}
In this section, the optimization problems for minimum energy consumption in multi-operator cellular communication are formulated for two different cases: the non-collaborative and collaborative cases. We derive the expressions of the energy consumption of operators and their operational profits that are used in the problems' objectives and constraints.
\subsection{Non-collaborative Case}
In the traditional case, i.e., the non-collaborative case, operators do not collaborate with each other for serving the users. Therefore, each operator only serves its own users despite the overlapping coverage regions with other operators. In order to save energy, we employ the BS sleeping strategy. Let $\boldsymbol{\epsilon}_l^{(u)}$ be a binary vector that indicates the states of the BSs belonging to operator $l$ during the period $\Delta t$. Its elements $\epsilon_{lj}^{(u)}$ indicate whether a BS $j$ of cellular operator $l$ is turned off or not as follows:
\begin{equation}
\epsilon_{lj}^{(u)}=\left\{
\begin{array}{l}
  1 \mbox{      if BS $j$ is turned on},\\
  0 \mbox{      if BS $j$ is turned off}.
\end{array}\right.
\end{equation}
The number of ones and the number of zeros in $\boldsymbol{\epsilon}_l^{(u)}$ indicate the number of active and inactive BSs, respectively. Thus, the energy consumption and the corresponding energy cost imposed to operator $l$ for the non-collaborative case, denoted by ${\mathcal E}_{l}^{(u)}$ and ${\mathcal \psi_{l}}^{(u)}$ respectively, are given as:
\begin{equation}
{\mathcal E}_{l}^{(u)}=\sum_{j=1}^{N_{\mathrm{BS}}^{(l)}}\epsilon^{(u)}_{lj} q_{lj}^{(u)},\mbox{ and }
{\mathcal \psi_{l}}^{(u)}=\pi_{l}{\mathcal E}_{l}^{(u)}.
\label{Energy_uncoop}
\end{equation}
The profit ${\mathcal P}_l^{(u)}$ of operator $l$ corresponding to its individual operation\footnote{The profit considered in this paper corresponds to the gain obtained due to the deployed BSs operation in the area of interest. This profit does not represent the total operator's profit which is complex to determine. When collaborating, the operator will share few information (some of them are known, such as energy and service prices, etc.) that do not affect their confidentiality.} in area $\mathcal{A}$ is expressed~as:
\begin{equation}
{\mathcal P}_l^{(u)}=\sum_{\sigma=1}^{\Sigma_l}N_U^{(l,\sigma)} p^{(l)}_\sigma+R_{\mathrm{op}}\left(N_U^{(l)}\right)-{\psi}_{l}^{(u)},
\label{profit_uncoop_eq}
\end{equation}
\normalsize
where $N_U^{(l,\sigma)}$ is evaluated as $N_U^{(l,\sigma)}=\sum_{j=1}^{N_{\text{BS}}^{(l)}}N_{A_{lj}}^{(l,\sigma,u)}$, $p^{(l)}_\sigma$ is the unitary cost of the service $\sigma$ paid by a user of cellular operator $l$, and $R_{\mathrm{op}}$ is a constant extra revenue due to subscriptions. Each operator wishes to minimize its energy consumption as follows:
\normalsize
\begin{align}
\small
&\underset{\boldsymbol{\boldsymbol{\epsilon}_{lj}^{(u)}}}{\text{Minimize}}
{\mathcal E}_l^{(u)}=\sum_{j=1}^{N_{\mathrm{BS}}^{(l)}}\epsilon^{(u)}_{lj} q_{lj}^{(u)}, \label{eq:optim}\\
&\text{Subject to:}\;
P_{\text{tx},j}^{(l,u)} \leq \bar{P}, \; \forall j = 1,\ldots, N_{\text{BS}}^{(l)}  , \forall l = 1,\ldots, N_{\text{op}}, \label{const11}\\
&\sum_{\sigma=1}^{\Sigma_l}N_{A_{lj}}^{(l,\sigma,u)} \leq \bar{K} , \; \forall j = 1,\ldots, N_{\text{BS}}^{(l)} , \forall l = 1,\ldots, N_{\text{op}}.  \label{const12}
\end{align}
\normalsize
This problem is constrained by (\ref{const11}) that ensures that all BSs radiate power within the available power budget $\bar{P}$, and \eqref{const12} which implies that each BS can only support a maximum of $\bar{K}$ users. \textcolor{black}{Note that, in~\eqref{eq:optim}, minimizing the total energy consumption is equivalent to minimizing the energy cost and hence, maximizing the total operational profit in the non-collaborative case given in~\eqref{profit_uncoop_eq}}. Finding the optimal solution of this problem, especially for large scale networks, is complicated since the decision variables correspond to large binary vectors of length $N_{\mathrm{BS}}^{(l)}$. We propose an iterative algorithm to solve this problem later in this paper. Once the optimization problem expressed in~\cref{eq:optim,const11,const12} is solved for each operator $l$ by finding the best solution $\boldsymbol{\epsilon}_l^{(u)*}$, we can determine the corresponding energy consumption $\mathcal{E}_{l}^{(u)*}$ using~\eqref{Energy_uncoop} and the maximum profit ${\mathcal P}_l^{(u)*}$ using~\eqref{profit_uncoop_eq}. These results are used as a benchmark for comparison in the simulations.

\subsection{Collaborative Case} \label{sec5}
In the collaborative case, the operators have the opportunity to further reduce their energy costs by turning off their lightly loaded BSs and using the infrastructure of other operators to serve their users. In return, these operators pay an extra charge for the shared operation, known as the \emph{roaming price}. Additionally, due to collaboration, the coverage region of each BS is smaller than the non-collaborative case as illustrated in Fig.~\ref{voronoi_fig} in Section~\ref{simulation_model_sec}. Therefore, the BSs of collaborating operators will consume less energy due to the smaller coverage region and consequently, lesser number of users served, while retaining the flexibility to turn off their redundant BSs. Hence, collaboration may lead to significant energy cost savings for the operators. Similar to~\eqref{Energy_uncoop}, the total energy consumption and the corresponding cost for operator $l$ in the collaborative case are given as:
\begin{equation}
{\mathcal E}_{l}^{(c)}=\sum_{j=1}^{N_{\mathrm{BS}}^{(l)}}\epsilon^{(c)}_{lj} q_{lj}^{(c)},\mbox{ and }
{\mathcal \psi_{l}}^{(c)}=\pi_l{\mathcal E}_{l}^{(c)}.
\label{Energy_coop}
\end{equation}
Notice that, in the collaborative case, the amount of energy $q_{lj}^{(c)}$ includes the transmit power $P^{(l,c)}_{\text{tx},j} $ used to serve the network $l$'s users in addition to other roamed users belonging to other operators. The profit of operator $l$ under the collaborative setting ${\mathcal P}_l^{(c)}$ is expressed as a function of the inter-operator roaming prices as follows:
\begin{align}
{\mathcal P}_l^{(c)}&=\sum_{\sigma=1}^{\Sigma_l}N_U^{(l,\sigma)} p^{(l)}_\sigma+\sum_{t=1\atop  t\neq l}^{{N_{\text{op}}}}\sum_{j=1}^{N_{\text{BS}}^{(l)}} p_{tl} N_{j}^{(t \rightarrow l)}\nonumber\\
& - \sum_{t=1\atop  t\neq l}^{{N_{\text{op}}}}\sum_{j=1}^{N_{\text{BS}}^{(t)}}p_{lt} N_{j}^{(l \rightarrow t)}+R_{\mathrm{op}}\left(N_U^{(l)}\right)-\psi_{l}^{(c)},
\label{profit_coop_eq}
\end{align}
where $p_{tl}$ is the roaming price paid by operator $t$ to operator $l$ and $N_{j}^{(t \rightarrow l)}$ denotes the number of roamed users of operator $t$ served by $j^{\text{th}}$ BS of operator $l$. We assume symmetric roaming prices, i.e., $p_{tl} = p_{lt}, \forall \ l,t = 1,\ldots,N_{\text{op}}, t \neq l$. In other words, we consider that any two operators in the network will agree on a fixed roaming price for serving each other's users. \textcolor{black}{This assumption is considered for four different reasons. (i) \textcolor{black}{It is more balanced and equitable that operators impose the same unitary roaming price to each other during collaboration} and hence, the collaboration decision depends only on the number of roamed users. (ii) The framework's objective is to ensure fossil fuel consumption reduction without degrading the operators' profits. Hence, the collaborative operators are focusing on establishing agreements to reach this green goal rather than competing against each other to maximize their profits. Therefore, their profits before collaboration are set as references for their collaboration decision. (iii) As it will be shown in the sequel, a system of linear inequalities will be formulated and thanks to this assumption, the number of variables will be less than the number of inequalities which ensures that the system does not become underdetermined. (iv) Finally, in this way, the complexity of the system is reduced as the number of unknown roaming prices is reduced by a factor of 2.} Let $\mathbf{p}_{r}$ be an $n \times 1$ vector containing the inter-operator roaming prices $p_{lt}$, where $n=\frac{N_{\text{op}}\left(N_{\text{op}}-1\right)}{2}$. \textcolor{black}{In practice, such cooperative behaviour can be enforced by an environmental regulator by imposing a penalty on the operators based on their energy consumption per subscriber}.

The objective is to identify the set of BSs of each operator that should be turned off and the inter-operator roaming prices that result in the minimum energy consumption for each operator. This problem can be formulated as follows:
\begin{align}
&\underset{\boldsymbol{\boldsymbol{\epsilon}^{(c)}}, \mathbf{p}_{r} }{\text{Minimize}} \,\,\,{\mathcal E}_l^{(c)} = \sum_{j=1}^{N_{\mathrm{BS}}^{(l)}}\epsilon^{(c)}_{lj} q_{lj}^{(c)}, \ \ l = 1,\ldots,N_{\text{op}}, \label{eq_collaborative}\\
&\text{Subject to:}\;
\mathcal{P}_{l}^{(c)} \geq \mathcal{P}_{l}^{(u)}, \; l = 1,\ldots, N_{\text{op}},  \label{const1_coop} \\
&P_{\text{tx},j}^{(l,c)} \leq \bar{P}, \; \forall j = 1,\ldots, N_{\text{BS}}^{(l)} , \forall l = 1,\ldots, N_{\text{op}},  \label{const2_coop} \\
&\sum_{\sigma=1}^{\Sigma_l}N_{A_{lj}}^{(l,\sigma,c)} \leq \bar{K} , \; \forall j = 1,\ldots, N_{\text{BS}}^{(l)} , \forall l = 1,\ldots, N_{\text{op}}, \label{const3_coop}
\end{align}
where $\boldsymbol{\epsilon}^{(c)} = [{\boldsymbol{\epsilon}_{1}^{(c)}}^{T},  {\boldsymbol{\epsilon}_{2}^{(c)}}^{T}, \ldots,
{\boldsymbol{\epsilon}_{N_{\text{op}}}^{(c)}}^{T}]^{T}$ is the aggregated binary vector of the BSs of all operators indicating their ON/OFF status and $(.)^{T}$ denotes the matrix transpose operation. The constraint in \eqref{const1_coop} ensures that the profit of all operators in the collaborative case exceeds their corresponding profits in the non-collaborative case. In other words, the operators will agree to collaborate only if collaboration leads to an increase in their profit. The constraints \eqref{const2_coop} and \eqref{const3_coop} ensure that the BS power budget and its capacity are not exceeded. \textcolor{black}{Hence, the collaboration decision is subject to two conditions: 1) the capacity and power budget of the host BSs are not exceeded,i.e., QoS is maintained (see equation \eqref{transmitPowereq}), and 2) both operators maintain profit levels that are at least equal to the ones before collaboration. The roaming prices $\mathbf{p}_{r}$ are introduced to compensate the cost of the extra energy consumed by the host operator to serve the offloaded users.}

\section{Roaming Price Optimization and BS Sleeping Strategy} \label{green_solution}
\begin{figure}[t]
  \centering
  \includegraphics[height=4.75in]{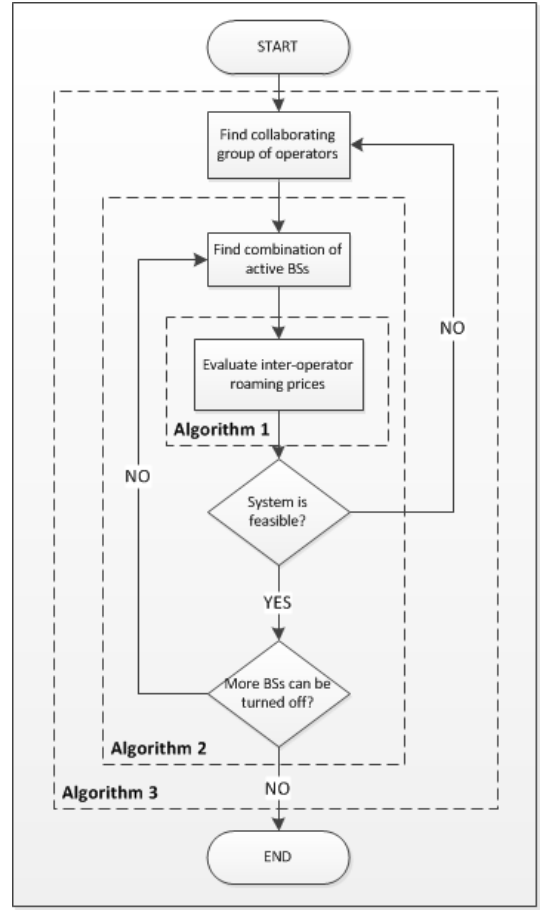}
  \caption{Top level activity diagram of the proposed virtualization framework.}\label{schema}
\end{figure}

The optimization problem in~\cref{eq_collaborative,const1_coop,const2_coop,const3_coop} is a combinatorial optimization problem where the objective is to jointly select the BSs that should be turned off as well as to find the roaming prices between operators in order to minimize the energy consumption of all operators. We will use a two-step process to solve this problem. In the first step, we assume that we have a given BS sleeping strategy characterized by  the set of active BSs of all operators $\boldsymbol{\epsilon}^{(c)}$. Given this $\boldsymbol{\epsilon}^{(c)}$, we optimize the inter-operator roaming prices for maximum profitability. In the second step, we propose an algorithm to determine the optimal BS status vector $\boldsymbol{\epsilon}^{(c)*}$. Both of these steps are performed iteratively to achieve the joint solution of the problem.  A top level activity diagram of the proposed methodology is provided in Fig.~\ref{schema}. Initially it is assumed that all operators are able to collaborate with each other. Assuming that all BSs are active, the inter-operator roaming prices are evaluated. If it turns out that the system is not feasible, i.e., collaboration is not possible between all operators, or further BSs cannot be turned off, the group of collaborating operators and/or the BS combination is updated. This procedure is repeated until, the final set of collaborating operators have been obtained with optimized active set of BSs as well as inter-operator roaming prices. \textcolor{black}{As mentioned in the introduction, the proposed green virtualization approach for operators collaboration is applied for long-term periods, e.g., 15 minutes, 1 hour, etc., during which the average statistics of the networks are almost static. Indeed, this solution can be part of a planning process involving infrastructure sharing ensuring green collaboration between operators. Based on previous networks' statistics, operators can decide the best moment at which the proposed virtualization approach is executed to determine the set of collaborative operators and the best BSs combination to be activated.}

In the following subsections, we develop Algorithm~\ref{Algorithm1} to optimize the inter-operator roaming prices among collaborating groups of operators. Algorithm~\ref{Algorithm2}, in Section~\ref{sec_ONOFF}, provides the procedure for optimizing the set of active BSs. Finally, Algorithm~\ref{Algorithm_grouping}, given in Section~\ref{sec_grouping}, describes the steps for identifying groups of collaborating operators. These algorithms are interconnected with each other as indicated in Fig.~\ref{schema}.

\subsection{Optimal Roaming Price Evaluation} \label{roaming_price_selection}
Given a particular active BS combination $\boldsymbol{\epsilon}^{(c)}$, the problem in~\cref{eq_collaborative,const1_coop,const2_coop,const3_coop} reduces to optimizing the roaming prices between operators for this set of active BSs since the constraints~\eqref{const2_coop} and~\eqref{const3_coop} only depend on $\boldsymbol{\epsilon}^{(c)}$. The reduced problem becomes a MOLPP and its solution is non-trivial since the objective of each operator is conflicting with the objective of other operators.

We employ an iterative method inspired from~\cite{MOLPP_iterative} that incorporates the idea of collaborative game theory~\cite{coalitional_game_theory} to solve the MOLPP. The operators are the players of the game and the inter-operator roaming price is the strategy selected by the players. The iterative algorithm is run until all players achieve a stable solution known as the Nash equilibrium~\cite{nash_equilibria}. The reduced problem can be formally stated as follows:
\normalsize
\begin{align}
&\underset{\mathbf{p}_{r} }{\text{Maximize}} \,\,{\mathcal P}_l^{(c)} = \sum_{\sigma=1}^{\Sigma_l}N_U^{(l,\sigma)} p^{(l)}_\sigma+\sum_{t=1\atop  t\neq l}^{{N_{\text{op}}}}\sum_{j=1}^{N_{\text{BS}}^{(l)}} p_{tl} N_{j}^{(t \rightarrow l)} \nonumber\\
&\hspace{1.5cm}- \sum_{t=1\atop  t\neq l}^{{N_{\text{op}}}}\sum_{j=1}^{N_{\text{BS}}^{(t)}}p_{lt} N_{j}^{(l \rightarrow t)} +R_{\mathrm{op}}\left(N_U^{(l)}\right)-\psi_{l}^{(c)}, \notag\\& \hspace{4cm}  \hfill{\forall l = 1,\ldots,N_{\text{op}}, }  \label{reduced_eq_collaborative} \\
&\text{Subject to:}\;\, \mathcal{P}_{l}^{(c)} \geq \mathcal{P}_{l}^{(u)}, \; \forall l = 1,\ldots, N_{\text{op}}.  \label{reduced_const1}
\end{align}
\normalsize
Each individual problem given in~\eqref{reduced_eq_collaborative}-\eqref{reduced_const1} has a unique optimum $\mathcal{P}_l^{(c)*}$ at $\mathbf{p}_{r}^{(l)*}$ for $l = 1, \ldots, N_{\text{op}}$, respectively, over the bounded set of constraints  $\mathcal{S}$, where $\mathcal{S}=\{\mathbf{p}_{r}\in \mathbb{R}^{n} : \mathcal{P}_{l}^{(c)} \geq \mathcal{P}_{l}^{(u)}, \forall l = 1,\ldots, N_{\text{op}} \}$. The objective of the problem is to achieve an equilibrium where all operators are able to achieve the maximum profit jointly at the same point, denoted by $\hat{\mathbf{p}}_{r}$. \textcolor{black}{From~\cref{reduced_eq_collaborative,reduced_const1}, it is clear that the optimal roaming price solution depends on the number of users that can be offloaded to the BSs of the other operator in case of collaboration as well as the non-collaborative profit}. The system of constraints in $\mathcal{S}$ can be easily written in the form of $\mathbf{A} \mathbf{p}_{r} \leq \mathbf{b}$, where $\mathbf{A}$ is a matrix of size $N_{\text{op}} \times n$ containing the constant coefficients of the system of linear inequalities corresponding to the number of roamed users from operator $t$ to $l$ and vice versa and $\mathbf{b}$ is a vector of size $N_{\text{op}} \times 1$ that contains the constant terms related to the service price and energy cost during the collaboration mode in addition to other term obtained from the non-collaborative profit expression given in \eqref{profit_uncoop_eq}. Each inequality determines a certain half-space while all the inequalities together determine a certain region in the $n$-dimensional space which is the intersection of a finite number of half spaces. If this linear system of inequalities admits a feasible solution, the operators can collaborate without degrading their profits. However, if the system is incompatible, collaboration among all operators is not possible. Nevertheless, there may exist disjoint groups of operators that are still able to collaborate among themselves. Therefore, the set of possibilities for the problem solution are listed below:

$\bullet$ The system of constraints is compatible: a feasible solution of the roaming prices $\mathbf{p}_{r}$ exists for the optimization problem.

$\bullet$ The system of constraints is infeasible: no solution exists for the roaming price and collaboration is not possible among any of the operators.

$\bullet$ The system of constraints is infeasible but disjoint subsets of the constraints provide a feasible solution space. In this case, different solutions exists for the subsets of inequalities of the linear system. Hence, collaboration is possible among operators corresponding to each subset. As an example, if there are five operators, it may be possible that operators 1 and 2 collaborate with each other while operators 3, 4, and 5 collaborate with each other while collaboration between the two groups is not possible. The optimal roaming prices for each group can then be computed independently.

$\bullet$ The system of constraints is infeasible but there are overlapping subsets of the constraints that provide a feasible solution space. In this case, a decision has to be made to partition the operators into separate groups and obtain a feasible solution for each group. As an example, if there are five operators, it may be possible that operators 1 and 2 can collaborate with each other while operators 4 and 5 can collaborate with each other. However, operator 3 is able to collaborate with both groups. Hence, operator 3 needs to be classified in any one of the collaborating groups and the optimal roaming prices for the two groups are obtained separately.

The groups of operators that can collaborate among themselves can be obtained using the algorithm defined in Section~\ref{groups}. Assuming that we have identified the groups of collaborating operators, we apply an iterative algorithm on each group to determine the optimal roaming prices among them. \textcolor{black}{We will describe here one iteration of the algorithm for brevity. The complete algorithm is provided in Algorithm~\ref{Algorithm1}. Let $d^{(l)}$ be the profit aspiration level of operator $l$, i.e., operator $l$ aspires to achieve $\mathcal{P}_l^{(c)} \geq d^{(l)}$. The set of desired roaming prices is defined as $\mathcal{D} = \{\mathbf{p}_{r} : \mathcal{P}_l^{(c)} \geq d^{(l)}, l = 1, \ldots , N_{\text{op}}\}$. As a starting point, we set the aspirations of all operators to the maximum level, i.e., $d^{(l)} = \mathcal{P}_l^{(c)*}$ for $l = 1,\ldots,N_{\text{op}} $. Since the maximum aspiration levels are achieved at different optimal points, the operators cannot collectively achieve their maximum desires at any one of these points. \textcolor{black}{Therefore, we define shifted desires $\lambda d^{(l)}$, which leads to a set of shifted desired roaming prices $\mathcal{D}_{\lambda} = \{\mathbf{p}_{r} : \lambda >0,  \mathcal{P}_l^{(c)} \geq \lambda d^{(l)} , l = 1, \ldots, N_{\text{op}} \}$ where $\lambda \in \mathbb R$}. If $\lambda$ is equal to $1$, it means that all operators are able to achieve their aspirations exactly. However, if $\lambda$ is less than or greater than $1$, it means that the operators are achieving less than or greater than their desired profit, respectively.} We want to maximize the value of $\lambda$ so that the highest possible fraction of the aspiration is achieved. To do this, we need to solve the following linear programming problem:
\begin{align}
&\underset{\mathbf{p}_{r} , \lambda}{\text{Maximize}} \ \ \ &&\lambda, \label{eq_lambda}\\
&\text{Subject to:}\; &&\mathcal{P}_{l}^{(c)} \geq \lambda d^{(l)} , \ l = 1, \ldots, N_{\text{op}},  \label{lambda_const1}\\
&&&\mathcal{P}_{l}^{(c)} \geq \mathcal{P}_{l}^{(u)}, \; l = 1,\ldots, N_{\text{op}}.  \label{lambda_const2}
\end{align}
\begin{algorithm}[t!]
\small
\caption{Iterative Algorithm for Solving the MOLPP}
\label{Algorithm1}
\begin{algorithmic}[1]
\STATE {t = 0; Let $\Psi = \left\{1,\ldots,N_{\text{op}}\right\}$ be the set of operators in the game and $d^{(l)}[t] \gets \mathcal{P}_l^{(c)*}$, $\forall l = 1,\ldots,N_{\text{op}}$.}
\STATE{$\delta^{(l)}_{\min} \gets \mathcal{P}_{l}^{(u)}$, $\delta^{(l)}_{\max} \gets \mathcal{P}_{l}^{(c)*}$.}
\REPEAT \label{loop_start}
\STATE $t \gets t + 1$.
\STATE Solve the optimization problem in \cref{eq_lambda,lambda_const1,lambda_const2} to obtain $\hat{\lambda}$ and $\hat{\mathbf{p}}_{r}$.
\STATE Determine the individual achievement of aspirations of each operator $\lambda^{(l)} \gets \frac{\hat{P}_{l}^{(c)}}{d^{(l)}}, \ l=1,\ldots,N_{\text{op}}$.
\STATE Calculate $l_{\text{max}} = \underset{l \in \Psi}{\text{argmax}} \ \lambda^{(l)} $ and $l_{\text{min}} = \underset{l \in \Psi}{\text{argmin}} \ \lambda^{(l)}$.
\IF {$\lambda^{(l_{\text{max}})} \geq 1$}
\STATE Remove $l_{\text{max}}$ from the set $\Psi$.
\ENDIF
\IF {$\Psi \neq \emptyset$}
\STATE Update minimum achieved aspiration using the dichotomic search algorithm as follows:
\IF {$\lambda^{(l_{\text{min}})} < 1 $}
\STATE $\delta_{\text{min}}^{(l_{\text{min}})} \gets d^{(l_{\text{min}})}[t]$,
\ELSE
\STATE $\delta_{\text{max}}^{(l_{\text{min}})} \gets d^{(l_{\text{min}})}[t]$.
\ENDIF
\STATE $d^{(l_{\text{min}})}[t+1] \gets \frac{\delta_{\text{max}}^{(l_{\text{min}})} + \delta_{\text{min}}^{(l_{\text{min}})} }{2}$.
\ENDIF
\UNTIL{$\hat{\lambda} \geq 1 \text{ or } \Psi = \emptyset$}. \label{loop_end}
\end{algorithmic}
\end{algorithm}
\normalsize
If $\mathcal{S} \cap D_{\lambda} \neq \emptyset$, then there exists a $\hat{\mathbf{p}}_{r}$ and $\hat{\lambda}$, that may or may not be unique. To measure the extent to which the aspiration level has been achieved by individual players after each iteration of the algorithm, we define the indicators $\lambda^{(l)} = \frac{\hat{P}_{c}^{(l)}}{d^{(l)}}, \ l= 1,\ldots,N_{\text{op}}$. $\hat{\lambda}$ corresponds to the minimum achievement of the aspirations of all operators. Since the operator corresponding to the least $\lambda^{(l)} < 1$ is not able to achieve its desired profit, it needs to decrease its aspiration $d^{(l)}$ in the next iteration. The updated $d^{(l)}$ can be decided using a dichotomic search algorithm (lines 12 - 18 in Algorithm~\ref{Algorithm1}). On the other hand, any player with $\lambda^{(l)} \geq 1$ can be removed from the game in the next iteration as it achieves its target. We repeat this process (lines \ref{loop_start} to \ref{loop_end}) successively until $\hat{\lambda}$ approaches $1$, i.e., all operators achieve their desired maximum profits. \textcolor{black}{The prices corresponding to this $\hat{\lambda}$ are the optimal roaming prices. Note that due to the nature of the objective and constraints, if there exists a feasible solution, Algorithm~\ref{Algorithm1} will always converge to obtain the optimal $\hat{\lambda}$ and the solution will be stable. However, if there is no feasible solution, then the collaboration will not take place, and is also a stable solution.}
\begin{algorithm}[t!]
\small
\caption{Iterative Algorithm for BS Sleeping Strategy}
\label{Algorithm2}
\begin{algorithmic}[1]
\STATE $t \gets 0$; Assume all BSs are activated, i.e., $\boldsymbol{\epsilon}[t]=\left[1,\cdots,1\right]$. \label{alg2:start}
\REPEAT
\STATE $t \gets t + 1$; $\Phi \gets \emptyset$.
\FOR {$j=1, \cdots, \sum_{\substack{l=1}}^{N_{\text{op}}}N_{\text{BS}}^{(l)}$}
\STATE Turn off BS $j$ if it is not already turned off.
\STATE Check the BS power budget and number of served users constraints as expressed in \eqref{const2_coop} and \eqref{const3_coop}, respectively, for all cells.
\IF {\eqref{const2_coop} and \eqref{const3_coop} are still satisfied for all cells}
\STATE Add $j$ to $\Phi$ (i.e., BS $j$ that belongs to the set of BSs that can be safely turned off) and compute $\hat{\mathcal{E}}_{l}^{(c)}[t,j]$ from~\eqref{Energy_coop} and $\hat{\mathcal{P}}_{l}^{(c)}[t,j]$ after solving the optimization problem in \cref{reduced_eq_collaborative,reduced_const1} for the given $\boldsymbol{\epsilon}[t,j]$ using \textbf{Algorithm~\ref{Algorithm1}}.
\ELSE
\STATE BS $j$ cannot be turned off.
\ENDIF
\ENDFOR
\STATE Find BS $\hat{j}\in \Phi$ such that, when it is turned off, the total energy consumption of the network is minimum: $\hat{j} \gets \underset{j\in \Phi}{\text{argmin}}\; \sum_{l=1}^{N_{\text{op}}} \hat{\mathcal{E}}_{l}^{(c)}[t,j]$
\STATE BS $\hat{j}$ is completely and safely eliminated. $\boldsymbol{\epsilon}[t,\hat{j}] \gets 0$.
\UNTIL{No more BS can be turned off}. \label{alg2:repeat_end}
\STATE $\mathcal{E}_{l}^{(c)}[t] \gets \hat{\mathcal{E}}_{l}^{(c)}[t,\hat{j}]$, $\mathcal{P}_{l}^{(c)}[t] \gets \hat{\mathcal{P}}_{l}^{(c)}[t,\hat{j}]$, $\boldsymbol{\epsilon}[t] \gets \boldsymbol{\epsilon}[t,\hat{j}]$ for all $l = 1,\ldots, N_{\text{op}}$.
\STATE $T \gets t$.
\WHILE{$\sum_{l=1}^{N_{\text{op}}} \mathcal{E}_{l}^{(c)}[T] \geq \sum_{l=1}^{N_{\text{op}}} \mathcal{E}_{l}^{(u)}$ $\mathbf{or}$  $\left(\sum_{l=1}^{N_{\text{op}}} \mathcal{E}_{l}^{(c)}[T] \leq \sum_{l=1}^{N_{\text{op}}} \mathcal{E}_{l}^{(u)} \text{ and } \mathcal{P}_{l}^{(c)}[T] \leq \mathcal{P}_{l}^{(u)} , l= 1, \ldots, N_{\text{op}}\right) $}
\STATE $T \gets T - 1$.
\STATE $\mathcal{E}_{l}^{(c)}[T] \gets \hat{\mathcal{E}}_{l}^{(c)}[T,\hat{j}]$, $\mathcal{P}_{l}^{(c)}[T] \gets \hat{\mathcal{P}}_{l}^{(c)}[T,\hat{j}]$, $\boldsymbol{\epsilon}[T] \gets \boldsymbol{\epsilon}[T,\hat{j}]$ for all $l = 1,\ldots, N_{\text{op}}$.
\ENDWHILE
\STATE $\boldsymbol{\epsilon}^{*} = \boldsymbol{\epsilon}[T]$
\end{algorithmic}
\end{algorithm}
\normalsize

\subsection{Active BS Set Optimization}
\label{sec_ONOFF}
In Section~\ref{roaming_price_selection}, we obtain the optimal values for inter-operator roaming prices assuming a given BS combination. \textcolor{black}{A BS sleeping strategy is used to ensure additional energy savings essentially during non-peak hours by turning off under-utilized BSs. Indeed, the green virtualization scheme has a rare chance to be applied for highly loaded or saturated networks since all BSs need to be kept on in order to maintain the required QoS level for each operator.} The optimal solution of the ON/OFF switching can be obtained using the exhaustive search algorithm where all possible combinations are tested. Although it achieves the best BSs combination, this method requires a high computational complexity, particularly when the number of BSs is very high, where $2^{\sum_{\substack{l=1}}^{N_{\text{op}}}N_{\text{BS}}^{(l)}}$ iterations are executed in order to reach the optimal solution. In our work presented in~\cite{TVTjournal1}, we proposed and compared deterministic and heuristic algorithms employed for the BS sleeping strategy for a single operator. Results show that the low complexity iterative algorithm, where one BS is eliminated at each iteration, with a computational complexity of $\approx\mathcal{O}\left(\left({\sum_{\substack{l=1}}^{N_{\text{op}}}N_{\text{BS}}^{(l)}}\right)^2\right)$ is able to achieve performance close to the evolutionary algorithms (e.g., genetic algorithm and particle swarm optimization approach) with significant gains in terms of computational time. Therefore, in this paper, we extend the iterative algorithm of \cite{TVTjournal1} to a multi-operator scenario to optimize the binary variables $\mathbf{\epsilon}^{(c)}$.

As indicated by the for loop (line 4 to 12), Algorithm 2 tests all the BSs and compute the corresponding utilities. The selected BS that will be turned off is the one that when turned off, the maximum possible energy saving is reached (line 13), its elimination does not provoke a roaming price infeasibility (line 8), the capacity and power budget constraints of all the other active BSs are not violated (line 6). At each iteration, only one BS is eliminated. We have added a brief paragraph in Section IV-B to clarify the process.

\textcolor{black}{The BS ON/OFF algorithm tests all the active BSs and computes the total energy consumption of the virtual network. At each iteration, the algorithm eliminates one BS which is the one that, when turned off, leads to the minimum energy consumption while allowing collaboration, i.e., it is possible to find common roaming prices using the procedure defined in Algorithm~\ref{Algorithm1} without violating the constraints~\cref{const2_coop,const3_coop}.} The BS elimination process is repeated until no further BSs can be turned off. At the end of this process, we revert back to check if either the total energy consumption or operator profit is improved as compared to the non-collaborative case or not by elimination in order to select the best BS combination. Note that this check is done after the BS elimination process, because it may happen that by turning off certain BSs, the energy consumption or operator profit gets worse than before collaboration. Hence, it may undesirably stop the elimination process. The complete procedure of the iterative algorithm for optimized BS sleeping strategy selection in collaborative operators is provided in Algorithm~\ref{Algorithm2}. The same approach (lines \ref{alg2:start} to \ref{alg2:repeat_end}) can be used to optimize the BS sleeping strategy $\epsilon^{(l)}, l= 1,\ldots, N_{\text{op}}$ in the traditional, i.e., non-collaborative case.

\label{groups}
\begin{algorithm}[t!]
\small
\caption{Algorithm for Identifying Collaborating Groups}
\label{Algorithm_grouping}
\begin{algorithmic}[1]
\STATE {Complete $\gets$ 0; Let $\mathbf{\Xi} = \{\Psi, E, \Pi\}$, where $\Psi = \left\{1,\ldots,N_{\text{op}}\right\} $ is the set of operators, $E = [\mathcal{E}_{1}^{(u)}, \mathcal{E}_{2}^{(u)}, \ldots, \mathcal{E}_{N_{\text{op}}}^{(u)}]$ is the set of non-collaborative energy consumption of the operators and $\Pi = [\mathcal{P}_{1}^{(u)}, \mathcal{P}_{2}^{(u)}, \ldots, \mathcal{P}_{N_{\text{op}}}^{(u)}]$ is the set of non-collaborative profit. Let $\Xi_{i} = \{ \Psi_{i} , E_{i}, \Pi_{i} \}$, where $i$ denotes the $i^{th}$ element in each set. }
\STATE Let $\mathbf{\xi} = \emptyset$ be the set containing the disjoint groups of operators that can collaborate with each other.
\WHILE{complete = 0,}
\STATE{Sort $\mathbf{\Xi}$ based on $E$ in decreasing order.}
\FOR{$i = 1$ to $|\mathbf{\Xi}|$}
\FOR{$j=i+1$ to $|\mathbf{\Xi}|$}
\STATE Find the set $s_{ij} = \{ \mathbf{p}_{r} \in \mathbb{R}^n : \mathcal{P}_{l}^{(c)} \geq \mathcal{P}_{l}^{(u)}, l = i,j \}$ assuming all BSs are active. If $s_{ij} \neq \emptyset$, compute the corresponding collaborative energy consumption $\mathcal{E}_{ij}^{(c)}$ and profit $\mathcal{P}_{ij}^{(c)}$ by solving the optimization problem in \cref{eq_collaborative,const1_coop,const2_coop,const3_coop} using \textbf{Algorithm~\ref{Algorithm2}}.
\ENDFOR
\STATE $s_{opt} = \{s_{ij} : \mathcal{E}_{ij}^{(c)} = \underset{\{j: s_{ij} \neq \emptyset\} }{\min} (\mathcal{E}_{ij}^{(c)}), \forall i,j \}$.
\IF{$s_{opt} \neq \emptyset $}
\STATE Combine Op. $i$ and Op. $j$ into a single virtual operator, denoted by Op. $ij$. Remove the $i^{th}$ and $j^{th}$ entry in $\mathbf{\Phi}$ and add a new entry for the virtual operator $ij$ i.e., $\mathbf{\Xi} =  \mathbf{\Xi} \backslash \{ \Xi_i,\Xi_j\} \cup \{\Xi_{ij}\}$, where $\Xi_{ij} = (ij, \mathcal{E}_{ij}^{(c)}, \mathcal{P}_{ij}^{(c)})$.
\STATE Let Op. $i$ denote the combined Op. $ij$.
\STATE $i \gets i-1$.
\ELSE
\STATE Op. $i$ cannot collaborate with any other operator. Remove Op. $i$ from the set $\mathbf{\Xi}$ i.e., $\mathbf{\Xi} = \mathbf{\Xi} \backslash \Xi_{i}$ and add $\Xi_{i}$ to the set $\mathbf{\xi}$.
\ENDIF
\ENDFOR
\IF{ $i = |\mathbf{\Xi}|$}
\STATE Complete $\gets$ 1.
\ENDIF
\ENDWHILE
\end{algorithmic}
\end{algorithm}
\normalsize

\subsection{Collaborating Operator Groups Identification} \label{sec_grouping}
As highlighted in Section~\ref{roaming_price_selection}, collaboration may not be possible among all the operators. However, sub-groups of operators might still be able to exploit collaboration to reduce energy consumption. We use the procedure defined by Algorithm~\ref{Algorithm_grouping} to identify the various groups of collaborating operators. The procedure begins with a set of all the operators along with their corresponding energy consumption and profits before collaboration, denoted by $\mathbf{\Xi}$. The first step is to sort the operators in descending order based on their energy consumption. This is done to ensure that we start the grouping with the operator that is most power greedy in order to reduce its energy consumption first. We check if it can collaborate with any of the other operators by determining if a feasible roaming price can be obtained between them. If the operators can collaborate, then the corresponding energy and profit after collaboration is computed after finding the optimal set of active BSs using Algorithm~\ref{Algorithm2} and the roaming price for the combined virtual operator. From all the collaboration possibilities, we select the operator group that leads to the minimum energy consumption after collaboration. The selected operators are removed from the set $\mathbf{\Xi}$ and replaced by a combined virtual operator. However, if collaboration is not possible with any of the other operators, the operator under test is added to a new set $\xi$, denoting the final disjoint groups of collaborating operators. This process is repeated until all the operators have been successfully added to the set $\xi$.
\begin{figure}[]
\begin{subfigure}[t]{.45\textwidth}
  \centering
    \includegraphics[width=.9\textwidth]{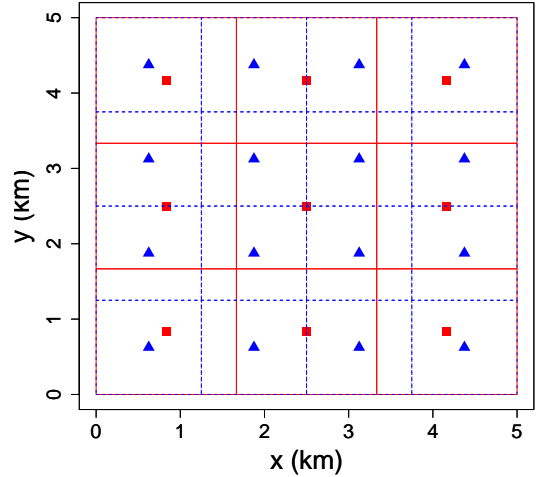}\vspace{-0.3cm}
    \caption{Non-collaborative mode. The solid lines grid represents the Voronoi cells for Op1 (red squares) while the dotted lines grid represents the Voronoi cells for Op2 (blue triangles).}
    \label{uncoop_fig}
  \end{subfigure}\hfill
  \begin{subfigure}[t]{.45\textwidth}
  \centering
    \includegraphics[width=.9\textwidth]{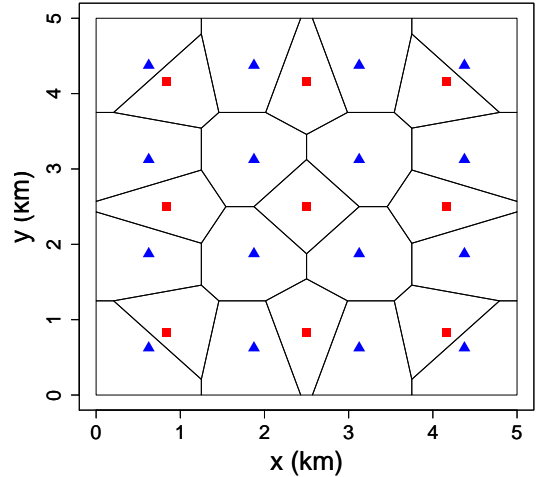}\vspace{-0.3cm}
    \caption{Collaborative mode. The BSs of Op1 (red squares) and Op2 (blue triangles) form a single virtual network.}
    \label{coop_fig}
  \end{subfigure}
	\vspace{-0.1cm}
\caption{Voronoi tessellation for $N_{\text{op}} = 2$ under non-collaborative and collaborative cases when all BSs are active.}
\label{voronoi_fig}
\end{figure}

\section{Results and Discussion}\label{sec7}
In this section, we investigate the performance of the proposed approach detailed in Section \ref{green_solution} for two and three operators. We begin by presenting the simulation model followed by the numerical results and discussion.
\subsection{Simulation Model} \label{simulation_model_sec}
For the sake of simple presentation, we initially consider $N_{\text{op}}=2$ operators denoted by Op1 and Op2, serving an area $\mathcal{A} = 5\times 5\, $km$^2$. Later, the case of $N_{\text{op}}=3$ operators is also considered for thorough investigation while maintaining interpretability of results. The BSs of Op1 and Op2 are deployed in $\mathcal{A}$ in the form of a well planned grid as shown by the squares and triangles in Fig.~\ref{voronoi_fig}(a). The number of BSs of Op1 is $N_{\mathrm{BS}}^{(1)}=9$ while the number of BSs of Op2 is $N_{\mathrm{BS}}^{(2)}=16$. We assume that the total number of users of Op1 are $N_{U}^{(1)} = 200$ and total number of users of Op2 $N_{U}^{(2)} = 150$ unless otherwise stated. \textcolor{black}{In the simulations, we focus mostly on the cases of relatively lightly loaded networks where establishing green collaboration mechanisms among operators is more frequent.} We assume that there is no inter-operator interference as the operators are using independent set of frequencies. All BSs have the same power model with the same maximum transmit power $46$ dBm, $a=7.84$, and $b=71.5$ W. We also assume that each BS can serve a maximum of $\bar{K}=50$ users. We set the path loss exponent $\eta=3.76$ and the path loss constant $K=-128.1$ dB. The minimum power level for detection is set as $P_{\text{min}} = -90$~dBm.

\begin{figure}[t!]
  \centering
  \includegraphics[width=8.5cm]{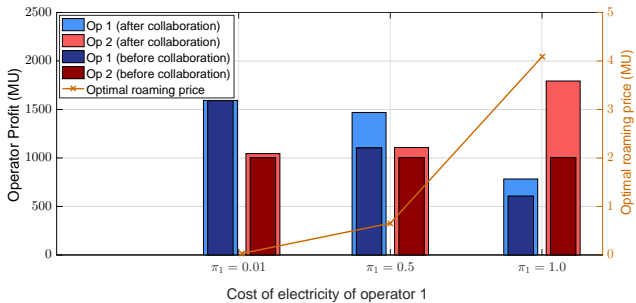}
  \caption{Operator profitability versus varying cost of electricity.}\label{fig2_new}
\end{figure}
\begin{figure}[t!]
  \centering
  \includegraphics[width=8.5cm]{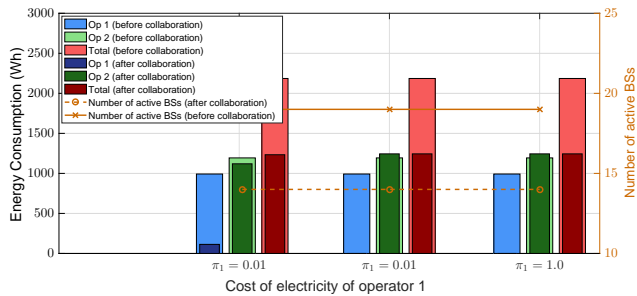}
  \caption{Energy consumption versus varying cost of electricity.}\label{fig1_new}
\end{figure}

Mobile operators are procuring energy either from electricity retailer that provides enough energy to cover the network operation or from locally generated renewable sources. We assume that the total amount of RE available to each operator is $0.45$ kWh that corresponds to the maximum amount of energy that is stored locally during \textcolor{black}{the operation time $\Delta t=1$ hour}. We set the unitary price of the fossil fuel to $\pi_{1}=\pi_{2}=0.5$ monetary units (MU). \textcolor{black}{For simplicity, we consider that each of the operators is providing a unique service $\Sigma_1=\Sigma_2=1$ with the following prices $p^{(1)}= 3$ MU and $p^{(2)} = 4$~MU.}

\subsection{Simulation Results}
The performance of operator collaboration under different electricity costs of the operators is studied in Fig.~\ref{fig2_new} and Fig.~\ref{fig1_new}. Particularly, we investigate three cases in which the fossil fuel costs for Op1 are less than, equal to, and greater than that of Op2. It can be observed in Fig.~\ref{fig2_new} that with the increase of Op1's energy cost, the profit of Op1 decreases successively. This motivates Op1 to turn off all its BSs and roam its users to Op2, i.e., collaborate at a fixed roaming price to reduce its energy consumption and increase profit. This can be observed in Fig.~\ref{fig1_new} from the fact that the energy consumption of Op1 after collaboration is $0$ kJ for the case of $\pi_{1} = 0.5$ and $\pi_{2} = 1$. It is clear that collaboration reduces the energy consumption of Op1 but increases the energy consumption of Op2. Since the roaming prices increase with the increase of energy cost of Op1, the profit of Op2 increases successively from $1044.8$ MU to $1793.4$ MU. Hence, the collaboration benefits both operators in terms of their individual energy consumption as well as profits. Most importantly, the overall energy consumption of the entire cellular network also decreases, e.g., for $\pi_{1} = 0.01$, the total energy consumption reduces from $0.22$ kWh to $1.23$ kWh, which helps reduce the carbon footprint of the network.

\begin{figure}[t!]
  \centering
  \includegraphics[width=8.5cm]{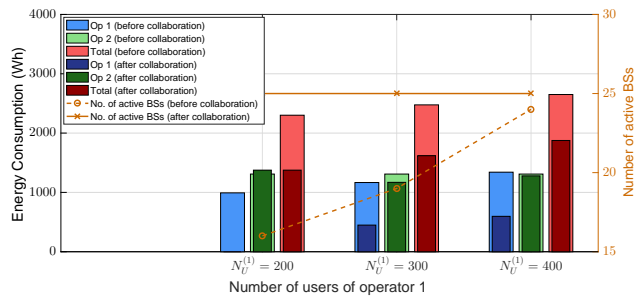}
  \caption{Energy consumption versus varying number of users of operator 1.}\label{fig3_new}\vspace{-0.1cm}
\end{figure}

\begin{figure}[t!]
  \centering
  \includegraphics[width=8.5cm]{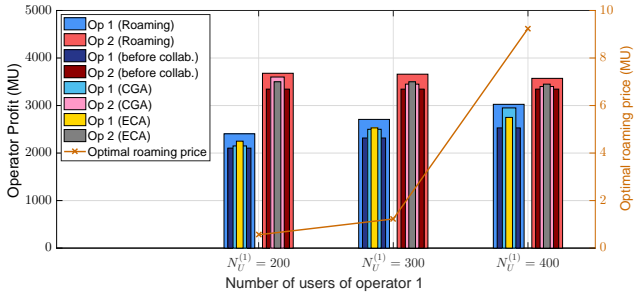}
  \caption{Operator profitability versus varying number of users of operator 1.}\label{fig4_new}
\end{figure}

In Fig.~\ref{fig3_new} and Fig.~\ref{fig4_new}, we study the impact of increasing the number of users on the collaborative gain of operators. Assuming that Op2 has a fixed number of users, i.e., $N_{U}^{(2)} = 500$, the number of users of Op1 are chosen to be $N_{U}^{(1)} = 200$, $300$, and $400$ successively. The cost of fossil fuels for both operators is kept the same, i.e., $\pi_{1}=\pi_{2}= 0.5$. It can be observed from Fig.~\ref{fig3_new} that increasing the number of users of Op1 increases its energy consumption from $0.99$ kWh to $1.2$ kWh and hence, the network's energy cost. Therefore, the incentive to collaborate increases. Since Op2 has more number of BSs and a capacity to accommodate external users, Op1 can turn off all its BSs, i.e., reduce its energy consumption to $0$ kWh, and roams its users to Op2 in exchange for a fixed roaming price when its number of users is low ($N_{U}^{(1)} = 200$). When $N_{U}^{(1)}$ increases, collaboration remains possible but Op1 is forced to keep active some of its BSs. Hence, it energy consumption after collaboration is almost equal to $0.45$ kWh and $0.6$ kWh for $N_{U}^{(1)} = 300$ and $N_{U}^{(1)} = 400$, respectively. It can be also noted that, under collaboration, the operators' profits increase but are always higher than the profits before collaboration in all cases as shown in Fig~\ref{fig4_new}. We also notice that roaming prices increase with the load of the network. Indeed, the host operators impose higher cost to serve more roamed users and compensate the higher extra energy consumption. \textcolor{black}{A comparison of the performance of the proposed approach with other fairness criterion such as the ECA and CGA~\cite{6347622} is also provided in Fig.~\ref{fig4_new}.}

\begin{figure}[t]
\begin{subfigure}[t]{.45\textwidth}
  \centering
    \includegraphics[width=0.95\textwidth]{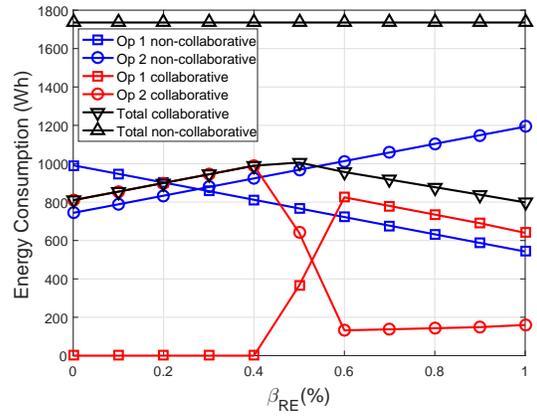}
    \caption{Energy consumption.}
    \label{energy_vs_green_fig}
  \end{subfigure}\hfill
  \begin{subfigure}[t]{.45\textwidth}
  \centering
    \includegraphics[width=0.95\textwidth]{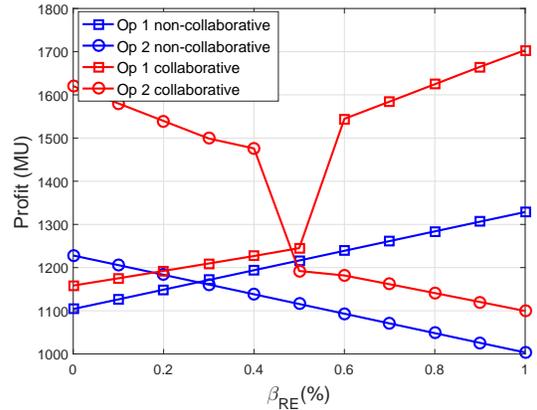}
    \caption{Operator profit.}
    \label{profit_vs_green_fig}
  \end{subfigure}
\caption{(a) Energy consumption and (b) operator profit versus the distribution of green energy between the operators.}
\label{energy_and_profit_fig}
\end{figure}

Next, we investigate the impact of generating RE by operators on the collaboration performance for $N_{U}^{(1)}=200$ and $N_{U}^{(2)}=150$. To do this, we introduce a parameter $\beta_{RE}$ that represents the percentage of green energy generated by Op1 while $100-\beta_{RE}$ corresponds to the percentage of green energy generated by Op2. In other words, if $\beta_{RE}=0\%$, then only Op2 makes use of green RE and vice versa. We assume here, for fairness, that both operators have the same maximum amount of green energy available, i.e., $0.450$ kWh, which is distributed equally among all BSs. Fig.~\ref{energy_vs_green_fig} shows the fossil fuel consumption of the operators in response to the percentage of green energy utilization. As Op1 becomes more dominant in the use of RE compared to Op2, its fossil fuel consumption decreases while the fossil fuel consumption of Op2 increases in the non-collaborative mode. Under the collaborative scenario, the BSs of Op1 are all turned off if it is not utilizing green energy. As Op1 starts to increase utilizing the locally generated green energy relative to Op2, the fossil fuel consumption of Op2 increases. As soon as Op1 becomes the dominant user of green energy, some of its BSs are turned on due to the availability of free energy. As a result, the energy consumption of Op1 sharply rises but keeps decreasing as we increase the usage of green energy. Since Op1 is now servicing users, the energy consumption of Op2 reduces sharply due to smaller number of served users.
\begin{figure}[t]
\begin{subfigure}[t]{.45\textwidth}
  \centering
    \includegraphics[width=0.95\textwidth]{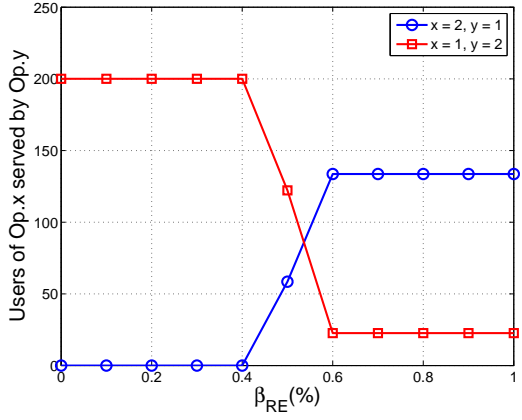}
    \caption{Roamed users.}
    \label{roamed_users_fig}
  \end{subfigure}\hfill
  \begin{subfigure}[t]{.45\textwidth}
  \centering
    \includegraphics[width=0.95\textwidth]{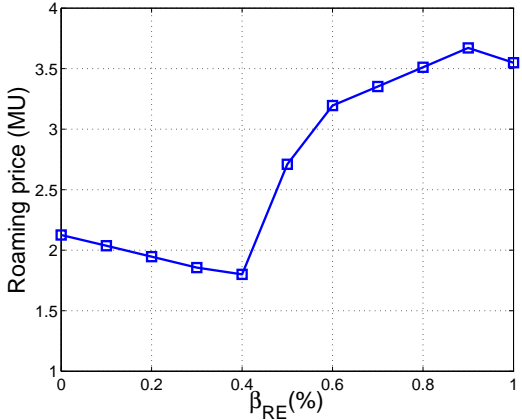}
    \caption{Roaming price.}\vspace{-0.2cm}
    \label{roaming_price_vs_green_fig}
  \end{subfigure}
\caption{(a) Number of roamed users and (b) roaming price versus the distribution of green energy between the operators.}
\label{roamed_user_and_price_fig}
\end{figure}

The profit of the operators versus the green energy distribution between them is shown in Fig.~\ref{profit_vs_green_fig}. The profit is directly linked to the fossil fuel consumption and the cost of energy paid by each operator as evident from~\eqref{Energy_uncoop},~\eqref{profit_uncoop_eq},~\eqref{Energy_coop}, and~\eqref{profit_coop_eq}. For fixed energy costs, increasing the green energy usage of Op1 increases its profit while reducing the profit of Op2. This can be attributed to the decrease and increase in fossil fuel consumption of Op1 and Op2, respectively, as shown in Fig.~\ref{energy_vs_green_fig}. It can be observed, however, that in all cases the profit of both operators after collaboration is higher than their individual profits in the non-collaborative case.

\begin{figure}[t]
\begin{subfigure}[t]{.45\textwidth}
  \centering
    \includegraphics[width=.9\textwidth]{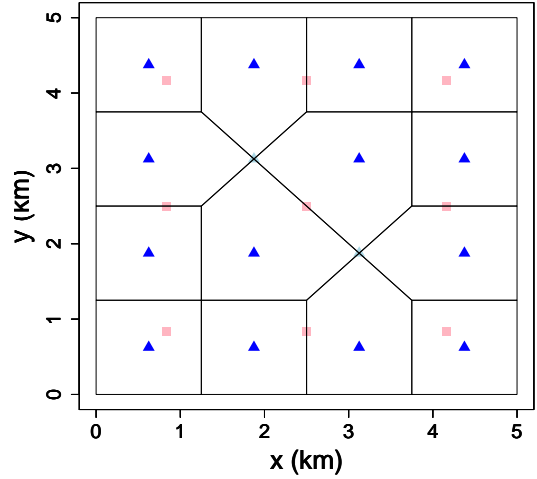}
    \caption{Set of active BSs under uniform user distribution.}
    \label{uniform_user_voronoi}
  \end{subfigure}\hfill
  \begin{subfigure}[t]{.45\textwidth}
  \centering
    \includegraphics[width=.9\textwidth]{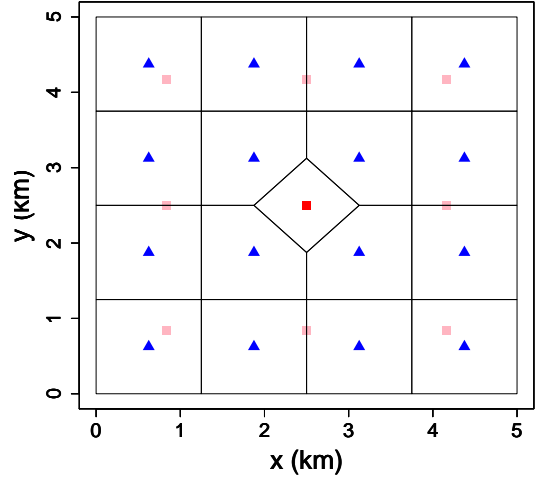}
    \caption{Set of active BSs under Gaussian user distribution centered at (2.5,2.5) with unit variance.}
    \label{gaussian_user_voronoi}
  \end{subfigure}
\caption{Optimized BS sleeping strategy under Uniform and Gaussian user distribution.}
\label{voronoi_fig2}
\end{figure}

In Fig.~\ref{roamed_users_fig}, we plot the number of users of Op1 served by Op2 and vice versa. As Op1 utilizes more green energy (i.e., $\beta_{RE}$ increases), its fossil fuel consumption, and subsequently the energy expenditure, decreases. Therefore, the system begins to transfer the users of Op2 to the BSs of Op1 so that both operators can increase their profits by reducing costs. The curves in Fig.~\ref{roamed_users_fig} are unbalanced because of the difference in the number of connected users and the number of available BSs for each operator. Finally, from Fig.~\ref{roaming_price_vs_green_fig}, we can notice that the roaming price is higher when Op1 is controlling the RE. Indeed, as the number of users of Op2 is lower, Op1 is forced to increase the roaming price in order to maximize its profit when collaborating while the inverse can be deduced for Op2.

The performance of the proposed framework has also been tested under different user distributions. Fig.~\ref{uniform_user_voronoi} shows the set of active BSs as well as the cell shapes if the users are uniformly distributed in the considered area. It is observed that all BSs of Op1 are switched off while only two BSs of Op2 are switched off to minimize the energy consumption. The cell sizes are large since some BSs are not active. However, in the case of Gaussian distribution of users centered at (2.5,2.5) with unit variance, i.e., users concentrated towards the center of the grid, the number of users in the center are very high. An optimal BS sleeping policy is forced to turn on more BSs in the center of the grid to serve the additional users as shown in Fig.~\ref{gaussian_user_voronoi}. This demonstrates that the proposed framework is adaptive with the spatial distribution of the users.

In practice, the roaming price cannot be varied instantaneously and dynamically for each combination of channel realizations in the network. It can have a pre-defined fixed average value for a given traffic density, or range of traffic densities in the network (e.g., there can be a price during the day corresponding to high density and another during the night corresponding to relatively lower density). This value can be set through collaboration agreements between mobile operators. The results derived in this paper are based on average values for the number of users and the transmit powers of the BSs. Hence, these results provide insights about the average roaming price that should be imposed between mobile operators for different traffic densities in order to ensure mutual benefit.

\begin{table}[]
\centering
\caption{Collaboration in the case of three operators}
\label{3optable}
\addtolength{\tabcolsep}{-0.1cm}
\resizebox{0.45\textwidth}{!}{%
\begin{tabular}{|c|c|c|c|c|c|}
\hline
\multirow{2}{*}{Groups}     & \multicolumn{5}{c|}{$N_{U}^{(1)}$ = 10}                                                                                                                                                                                                                                                                            \\ \cline{2-6}
                            & \begin{tabular}[c]{@{}c@{}}Collaboration\\ possibility\end{tabular} & \begin{tabular}[c]{@{}c@{}}Total energy\\ consumption\\ before {[}after{]}\\ collaboration (Wh)\end{tabular} & \begin{tabular}[c]{@{}c@{}}Number of\\ active BSs\end{tabular} & \begin{tabular}[c]{@{}c@{}}Roaming\\ price (MU)\end{tabular} & \begin{tabular}[c]{@{}c@{}}Profit\\ Increase\\Op. 1, Op. 2, Op. 3 \\ (\%)\end{tabular}\\ \hline
Op. 1 and Op. 2                                                   & Yes                                                                 & \begin{tabular}[c]{@{}c@{}}2694.4\\ {[}2054.8{]}\end{tabular}                                           & 0 + 10 + 11                                                    & 119.31              &             1290, 81, 0                       \\ \hline
Op. 1 and Op. 3                                                   & No                                                                  & \begin{tabular}[c]{@{}c@{}}N/A\\ {[}N/A{]}\end{tabular}                                                                                                     & 9 + 10 + 11                                                    & N/A                          &            N/A               \\ \hline
Op. 2 and Op. 3                                                   & \textbf{Yes}                                                                 & \begin{tabular}[c]{@{}c@{}}\textbf{2694.4}\\ {[}\textbf{1877.8}{]}\end{tabular}                                           & \textbf{9 + 12 + 3}                                                     & \textbf{29.82}                       &           0, 5.9, 0.02                 \\ \hline
\begin{tabular}[c]{@{}c@{}}Op. 1, Op. 2,\\ and Op. 3\end{tabular} & No                                                                  & \begin{tabular}[c]{@{}c@{}}N/A\\ {[}N/A{]}\end{tabular}                                                                                                    & 9 + 10 + 11                                                    & N/A                            &              N/A           \\ \hline
\end{tabular}}
\\ \vspace{0.2in}
\resizebox{0.45\textwidth}{!}{%
\begin{tabular}{|c|c|c|c|c|c|}
\hline
\multirow{2}{*}{Groups}                                           & \multicolumn{5}{c|}{$N_{U}^{(1)}$ = 100}                                                                                                                                                                                                                                                                            \\ \cline{2-6}
                                                                  & \begin{tabular}[c]{@{}c@{}}Collaboration\\ possibility\end{tabular} & \begin{tabular}[c]{@{}c@{}}Total energy\\ consumption\\ before {[}after{]}\\ collaboration (Wh)\end{tabular} & \begin{tabular}[c]{@{}c@{}}Number of\\ active BSs\end{tabular} & \begin{tabular}[c]{@{}c@{}}Roaming\\ price (MU)\end{tabular} & \begin{tabular}[c]{@{}c@{}}Profit\\ Increase\\Op. 1, Op. 2, Op. 3 \\ (\%)\end{tabular} \\ \hline
Op. 1 and Op. 2                                                   & Yes                                                                 & \begin{tabular}[c]{@{}c@{}}2851.3\\ {[}2228.9{]}\end{tabular}                                           & 0 + 12 + 11                                                    & 15.68                                            &   86.2, 89.3, 0   \\ \hline
Op. 1 and Op. 3                                                   & Yes                                                                  & \begin{tabular}[c]{@{}c@{}}2851.3\\ {[}2160.9{]}\end{tabular}                                            & 0 + 10 + 13                                                    & 15.57                                            &   87.9, 11.8, 69.1     \\ \hline
Op. 2 and Op. 3                                                   & \textbf{Yes}                                                                 & \begin{tabular}[c]{@{}c@{}}\textbf{2851.3}\\ {[}\textbf{2034.8}{]}\end{tabular}                                           & \textbf{9 + 12 + 3}                                                     & \textbf{29.82}                                  &        0, 8.9, 0         \\ \hline
\begin{tabular}[c]{@{}c@{}}Op. 1, Op. 2,\\ and Op. 3\end{tabular} & No                                                                  & \begin{tabular}[c]{@{}c@{}}N/A\\ {[}N/A{]}\end{tabular}                                                                                                     & 9 + 10 + 11                                                    & N/A                           &             N/A             \\ \hline
\end{tabular}
}\\ \vspace{0.2in}
\resizebox{0.45\textwidth}{!}{%
\begin{tabular}{|c|c|c|c|c|c|}
\hline
\multirow{2}{*}{Groups}                                           & \multicolumn{5}{c|}{$N_{U}^{(1)}$ = 200}                                                                                                                                                                                                                                                                            \\ \cline{2-6}
                                                                  & \begin{tabular}[c]{@{}c@{}}Collaboration\\ possibility\end{tabular} & \begin{tabular}[c]{@{}c@{}}Total energy\\ consumption\\ before {[}after{]}\\ collaboration (Wh)\end{tabular} & \begin{tabular}[c]{@{}c@{}}Number of\\ active BSs\end{tabular} & \begin{tabular}[c]{@{}c@{}}Roaming\\ price (MU)\end{tabular} & \begin{tabular}[c]{@{}c@{}}Profit\\ Increase\\Op. 1, Op. 2, Op. 3 \\ (\%)\end{tabular}\\ \hline
Op. 1 and Op. 2                                                   & Yes                                                                 & \begin{tabular}[c]{@{}c@{}}3025.7\\ {[}2314.7{]}\end{tabular}                                           & 0 + 14 + 11                                                    & 9.8562                        &     22.4, 130.4, 2.14                     \\ \hline
Op. 1 and Op. 3                                                   & Yes                                                                  & \begin{tabular}[c]{@{}c@{}}3025.7\\ {[}2231.5{]}\end{tabular}                                                                                                     & 0 + 10 + 15                                                    & 9.7409                                &         13.02, 12.3, 85.8            \\ \hline
Op. 2 and Op. 3                                                   & Yes                                                                 & \begin{tabular}[c]{@{}c@{}}3025.7\\ {[}2209.1{]}\end{tabular}                                           & 9 + 12 + 3                                                     & 29.82                          &      0, 4.3, 0                   \\ \hline
\begin{tabular}[c]{@{}c@{}}Op. 1, Op. 2,\\ and Op. 3\end{tabular} & \textbf{Yes}                                                                  & \begin{tabular}[c]{@{}c@{}}\textbf{3025.7}\\ {[}\textbf{1334.8}{]}\end{tabular}                                                                                                     & \textbf{0 + 10 + 6}                                                    &         \begin{tabular}[c]{@{}c@{}}$\mathbf{p_{12} = 0.13}$,\\ $\mathbf{p_{13} = 0.47}$, \\ $\mathbf{p_{23} = 0}$\end{tabular}                            &        172, 15.5, 0       \\ \hline
\end{tabular}}
\vspace{-0in}
\end{table}

In Table~\ref{3optable}, we investigate the case of collaboration among three operators, i.e., $N_{\text{op}} = 3$. Op.1 and Op.2 are the same as in the previous simulations, while a new operator denoted by Op.3 is introduced with $N_{\text{BS}}^{(3)} = 16$. Specifically, the aim is to study the operation of Algorithm~\ref{Algorithm_grouping}, which helps in identifying collaborating groups of operators. For three operators, the possible collaborating groups can be as follows; Op.1 and Op.2 collaborate leaving out Op.3, Op.1 and Op.3 collaborate leaving out Op.2, Op.2 and Op.3 collaborate leaving out Op.1, or all three operators collaborate. We use the following simulation parameters for the results in Table~\ref{3optable}: $N_{U}^{(2)} = 100$, $N_{U}^{(3)} = 200$, $p^{(1)} = 8$ MU, $p^{(2)} = 1$ MU, $p^{(3)} = 2$ MU, $\pi_{1} = 2.5$ MU, $\pi_{2} = 0.5$ MU, $\pi_{3} = 0.1$ MU. The collaboration decisions are observed against increasing average number of users of Op.1 $N_{U}^{(1)}$. For low number of users, i.e., $N_{U}^{(1)}=10$, it is shown that collaboration is possible between Op.1 and Op.2 and also between Op.2 and Op.3 with energy saving of 24\% and 30\%, respectively. According to Algorithm~\ref{Algorithm_grouping}, the combination that achieves the lowest total energy consumption is selected. Hence, the combination of Op.2 and Op.3, highlighted in bold in the table, is selected for collaboration while Op.1 operates independently. Collaboration in other groups is not possible since there is no feasible roaming price.

As the number of users of Op. 1 increases to $N_{U}^{(1)} = 100$, collaboration becomes possible between any two operators while the simultaneous collaboration among all three operators is not feasible due to incompatible roaming prices. Again, we compute the total energy consumption for all possible groups of operators. The collaboration between Op.2 and Op.3 leads to maximum energy saving, i.e., 29\%. Further increase in the number of users of Op.1 to $N_{U}^{(1)} = 200$ provides more degrees of freedom for collaboration as all operators can collaborate simultaneously. The joint collaboration between Op.1, Op.2, and Op.3 leads to the maximum energy savings, i.e., 56\% since it allows the maximum number of BSs to be turned off.

The number of active BSs and the optimal roaming prices for the collaborating groups of operators are also provided in Table~\ref{3optable}. It can be observed, in general, that the number of active BSs increase with $N_{U}^{(1)}$ as more BSs are required to be turned on to serve additional users. On the other hand, the number of active BSs are reduced as a result of collaboration between operators. As an example, for the case of $N_{U}^{(1)} = 200$, when only Op.1 and Op.3 collaborate, the number of active BSs are $25$ ($0+10+15$) as compared to the case of $N_{U}^{(1)} = 10$ in which Op.1 and Op.3 cannot collaborate and the number of active BSs are 30 ($9+10+11$). Regarding the roaming prices, it is observed that they are decreasing with the increase of the number of users $N_{U}^{(1)}$. Indeed, as the number of roamed users increase, the roaming prices rise to cover the additional energy costs incurred by the serving operator. Secondly, the roaming prices also decrease as more operators collaborate together. This is due to the fact that the roamed users are shared among many operators instead of one and hence, the roaming prices are relaxed. For example, the inter-operator roaming prices when all three operators collaborate, is lower as compared to the case when any two operators collaborate for $N_{U}^{(1)}=200$.

\subsection{Discussions}
The results show that green collaboration between multiple operators can be beneficial and may lead to a win-win situation for all parties. Operators can increase their revenues from roamed users, while cutting their operating costs by significantly reducing their energy consumption. The selection of suitable roaming prices will allow the serving operator to generate revenue from roamed users, whereas the original operator will be able to save energy costs by turning off redundant BSs. In addition, the whole collaboration process is environment-friendly.

\textcolor{black}{Throughout the operation of the network, the roles of the operators will be reversed depending on the network dynamics. Hence, each operator will, at certain times (and/or at different locations), save energy by switching off some of its BSs, while at other times and/or locations, it will be serving the roamed users of other operators. Therefore, the operators need to pre-identify where and when virtualization can be applied. The proposed approach is applied offline at a certain geographical subarea where the number of BSs and the user statistics are pre-known or efficiently estimated, since each operator generally knows its network statistics. The same approach can be applied to other subareas and hence, multiple roaming prices are determined. Final roaming prices can be decided depending on the agreement signed by the collaborative mobile operators. For example, they can be dynamic and vary from a subarea to another depending on its characteristics or a final roaming price can be determined after evaluating all roaming prices at each subarea. Hence, it can corresponds to a weighted function of these local roaming prices. Then, if collaboration is decided during certain periods, the pre-agreed roaming prices are used online to calculate the dues of each operator towards the others.}

\textcolor{black}{When the operators are collaborating, the infrastructure of their respective networks will be shared and thus, it will act as the infrastructure of a single bigger network. The network access information will be shared accordingly. Thus, an active BS, regardless to which operator it belongs, will appear to subscribers of operator A as a BS of operator A, and to users of operator B as a BS of operator B. This can happen by letting the BS transmit, on two different channels, the signaling information corresponding to operator A and the signaling information corresponding to operator B so that users of both operators can identify it. In this case, this single BS will be acting as two virtual co-located BSs, one for operator A and another for operator B. Hence, if collaboration exists between the two operators, the users will be handed over seamlessly and transparently between BSs.}

In practice, this multi-operator collaboration is in line with the active research area of network function virtualization (NFV) and service orchestration. However, the main challenge is in pricing and billing issues. Operators need to find the best roaming pricing strategy that can lead to benefits for all involved parties. This requires an assessment of the value of the savings obtained by switching a certain BS off and of the costs incurred by the new serving operator to serve users of other operators. Once studies are made to estimate these values, suitable billing agreements can be signed between concerned operators. The whole process remains transparent to users, who will pay their bills to their initial operator and will receive their expected QoS seamlessly across the networks of the collaborating operators.

\section{Conclusions} \label{sec9}
In this paper, we proposed a green virtualization framework for collaboration of multiple cellular operators to achieve energy efficiency. The BS sleeping strategy is employed to switch off redundant BSs from the virtual network that is formed by unifying the radio access infrastructure of all operators. The users associated to the turned off BSs are off-loaded to other active BSs. Roaming prices are charged by operators to serve the users of other operators through their infrastructure. An optimization framework for identifying the active BS combination that minimizes the energy consumption and the set of inter-operator roaming prices that maximizes operator profits is formulated. The joint optimization is performed by solving a multi-objective linear programming problem using a game theoretic iterative algorithm. Moreover, an algorithm is proposed that helps in identifying the subset of operators that can collaborate with each other in case the collaboration among all operators is not possible due to profitability, capacity, or power constraints. It is shown that considerable energy can be saved under the proposed virtualization as compared to standalone operators. In view of the demonstrated benefits in terms of energy and profitability, it is recommended that the telecommunication leaders and regulators discuss and focus on such approaches for possible implementation in next generation of cellular networks.

\bibliographystyle{ieeetran}
\bibliography{bibpaper1}

\begin{thebibliography}{10}
\providecommand{\url}[1]{#1}
\csname url@samestyle\endcsname
\providecommand{\newblock}{\relax}
\providecommand{\bibinfo}[2]{#2}
\providecommand{\BIBentrySTDinterwordspacing}{\spaceskip=0pt\relax}
\providecommand{\BIBentryALTinterwordstretchfactor}{4}
\providecommand{\BIBentryALTinterwordspacing}{\spaceskip=\fontdimen2\font plus
\BIBentryALTinterwordstretchfactor\fontdimen3\font minus
  \fontdimen4\font\relax}
\providecommand{\BIBforeignlanguage}[2]{{%
\expandafter\ifx\csname l@#1\endcsname\relax
\typeout{** WARNING: IEEEtran.bst: No hyphenation pattern has been}%
\typeout{** loaded for the language `#1'. Using the pattern for}%
\typeout{** the default language instead.}%
\else
\language=\csname l@#1\endcsname
\fi
#2}}
\providecommand{\BIBdecl}{\relax}
\BIBdecl

\bibitem{cisco_report}
``Cisco visual networking index: Global mobile data traffic forecast update
  2014-2019,'' White Paper, Feb. 2015.

\bibitem{weo}
{International Energy Agency}, ``World energy outlook 2014,'' Nov. 2014.

\bibitem{ericsson_report}
A.~Fehske, G.~Fettweis, J.~Malmodin, and G.~Biczok, ``The global footprint of
  mobile communications: The ecological and economic perspective,'' \emph{IEEE
  Commun. Mag.}, vol. 49, no. 8, pp. 55--62, Aug. 2011.

\bibitem{7570259}
H.~Ghazzai, E.~Yaacoub, A.~Kadri, H.~Yanikomeroglu, and M.~S. Alouini,
  ``Next-generation environment-aware cellular networks: Modern green
  techniques and implementation challenges,'' \emph{IEEE Access}, vol.~4, pp.
  5010--5029, Sept. 2016.

\bibitem{energy_efficiency_need}
E.~Oh, B.~Krishnamachari, X.~Liu, and Z.~Niu, ``Toward dynamic energy-efficient
  operation of cellular network infrastructure,'' \emph{IEEE Commun. Mag.},
  vol.~49, no.~6, pp. 56--61, June 2011.

\bibitem{manifesto}
``{Mobile\rq s Green Manifesto 2012},'' Groupe Speciale Mobile Association
  (GSMA), Tech. Rep., 2012.

\bibitem{green_survey}
A.~Bianzino, C.~Chaudet, D.~Rossi, and J.~Rougier, ``A survey of green
  networking research,'' \emph{IEEE Commun. Surveys Tuts.}, vol. 14, no. 1, pp.
  3--20, First Quarter 2012.

\bibitem{green_survey_2}
Z.~Hasan, H.~Boostanimehr, and V.~Bhargava, ``Green cellular networks: A
  survey, some research issues and challenges,'' \emph{IEEE Commun. Surveys
  Tuts.}, vol. 13, no. 4, pp. 524--540, Second Quarter 2011.

\bibitem{cell_zooming}
Z.~Niu, Y.~Wu, J.~Gong, and Z.~Yang, ``Cell zooming for cost-efficient green
  cellular networks,'' \emph{IEEE Commun. Mag.}, vol.~48, no.~11, pp. 74--79,
  Nov. 2010.

\bibitem{network_virtualization}
C.~Liang and F.~Yu, ``Wireless network virtualization: A survey, some research
  issues and challenges,'' \emph{IEEE Commun. Surveys Tuts.}, vol.~17, no.~1,
  pp. 358--380, Aug. 2015.

\bibitem{infrastructure_sharing_report}
``Mobile infrastructure sharing,'' Groupe Speciale Mobile Association (GSMA),
  Tech. Rep., 2012.

\bibitem{infrastructure_sharing}
A.~Antonopoulos, E.~Kartsakli, A.~Bousia, L.~Alonso, and C.~Verikoukis,
  ``Energy-efficient infrastructure sharing in multi-operator mobile
  networks,'' \emph{IEEE Commun. Mag.}, vol.~53, no.~5, pp. 242--249, May 2015.

\bibitem{network_sharing}
M.~Marsan and M.~Meo, ``Network sharing and its energy benefits: A study of
  european mobile network operators,'' in \emph{IEEE Global Commun. Conf.
  ({Globecom} 2013)}, Atlanta, GA, USA, Dec. 2013.

\bibitem{7294664}
P.~D. Francesco, F.~Malandrino, T.~K. Forde, and L.~A. DaSilva, ``A sharing and
  competition-aware framework for cellular network evolution planning,''
  \emph{IEEE Trans. Cogn. Commun. Netw.}, vol.~1, no.~2, pp. 230--243, June
  2015.

\bibitem{greedy_ref_3}
L.~Mashayekhy, M.~M. Nejad, and D.~Grosu, ``Cloud federations in the sky:
  Formation game and mechanism,'' \emph{IEEE Trans. Cloud Comput.}, vol.~3,
  no.~1, Jan. 2015.

\bibitem{greedY_ref_2}
L.~Militano, A.~Orsino, G.~Araniti, A.~Molinaro, and A.~Iera, ``A constrained
  coalition formation game for multihop {D2D} content uploading,'' \emph{IEEE
  Trans. Wireless Commun.}, vol.~15, no.~3, Mar. 2016.

\bibitem{7056465}
H.~Ghazzai, E.~Yaacoub, M.~S. Alouini, Z.~Dawy, and A.~Abu-Dayya, ``Optimized
  {LTE} cell planning with varying spatial and temporal user densities,''
  \emph{IEEE Trans. on Veh. Tech.}, vol.~65, no.~3, pp. 1575--1589, Mar. 2016.

\bibitem{deployment1}
F.~Richter, A.~Fehske, and G.~Fettweis, ``Energy efficiency aspects of base
  station deployment strategies for cellular networks,'' in \emph{IEEE Veh.
  Technol. Conf. Fall (VTC-Fall 2009)}, Anchorage, AK, USA, Sept. 2009.

\bibitem{range_adaptation}
S.~Luo, R.~Zhang, and T.~J. Lim, ``Optimal power and range adaptation for green
  broadcasting,'' \emph{IEEE Trans. Wireless Commun.}, vol.~12, no.~9, pp.
  4592--4603, Sept. 2013.

\bibitem{greedy_Bousia}
A.~Bousia, A.~Antonopoulos, L.~Alonso, and C.~Verikoukis, ``Distance-aware base
  station sleeping algorithm in {LTE-Advanced},'' in \emph{IEEE Intl. Conf.
  Commun. (ICC 2012)}, Jun. 2012, pp. 1347--1351.

\bibitem{6489498}
E.~Oh, K.~Son, and B.~Krishnamachari, ``Dynamic base station switching-{On/Off}
  strategies for green cellular networks,'' \emph{IEEE Trans. Wireless
  Commun.}, vol. 12, no. 5, pp. 2126--2136, May 2013.

\bibitem{BS_sleeping_12}
J.~Wu, S.~Zhou, and Z.~Niu, ``Traffic-aware base station sleeping control and
  power matching for energy-delay tradeoffs in green cellular networks,''
  \emph{IEEE Trans. Wireless Commun.}, vol.~12, no.~8, pp. 4196--4209, Aug.
  2013.

\bibitem{BS_sleeping_11}
J.~Zheng, Y.~Cai, X.~Chen, R.~Li, and H.~Zhang, ``Optimal base station sleeping
  in green cellular networks: A distributed cooperative framework based on game
  theory,'' \emph{IEEE Trans. Wireless Commun.}, vol.~14, no.~8, pp.
  4391--4406, Aug. 2015.

\bibitem{small_cell_green}
Y.~S. Soh, T.~Quek, M.~Kountouris, and H.~Shin, ``Energy efficient
  heterogeneous cellular networks,'' \emph{IEEE J. on Sel. Areas in Commun.},
  vol.~31, no.~5, pp. 840--850, May 2013.

\bibitem{coop2}
M.~A. Marsan and M.~Meo, ``Energy efficient wireless internet access with
  cooperative cellular networks,'' \emph{Elsevier Computer Networks}, vol.~55,
  no.~2, pp. 386 -- 398, Feb. 2011.

\bibitem{coop3}
D.-E. Meddour, T.~Rasheed, and Y.~Gourhant, ``On the role of infrastructure
  sharing for mobile network operators in emerging markets,'' \emph{Elsevier
  Computer Networks}, vol.~55, no.~7, pp. 1576--1591, May 2011.

\bibitem{coop4}
A.~Bousia, E.~Kartsakli, A.~Antonopoulos, L.~Alonso, and C.~Verikoukis, ``Game
  theoretic approach for switching off base stations in multi-operator
  environments,'' in \emph{IEEE Int. Conf. on Commun. (ICC 2013)}, Dresden,
  Germany, June 2013.

\bibitem{7417231}
M.~Oikonomakou, A.~Antonopoulos, L.~Alonso, and C.~Verikoukis, ``Cooperative
  base station switching off in multi-operator shared heterogeneous network,''
  in \emph{IEEE Global Commun. Conf. (Globecom 2015)}, Dec. 2015, pp. 1--6.

\bibitem{6347622}
L.~Militano, A.~Molinaro, A.~Iera, and A.~Petkovics, ``Introducing fairness in
  cooperation among green mobile network operators,'' in \emph{Int. Conf. on
  Software, Telecommun. and Comput. Netw. (SoftCOM 2012)}, Split, Croatia,
  Sept. 2012.

\bibitem{renewable}
H.~Al~Haj~Hassan, L.~Nuaymi, and A.~Pelov, ``Renewable energy in cellular
  networks: A survey,'' in \emph{IEEE Online Conf. on Green Commun. (GreenCom
  2013)}, Oct. 2013.

\bibitem{BS_sleeping_renewable}
J.~Gong, J.~Thompson, S.~Zhou, and Z.~Niu, ``Base station sleeping and resource
  allocation in renewable energy powered cellular networks,'' \emph{IEEE Trans.
  Commun.}, vol.~62, no.~11, pp. 3801--3813, Nov. 2014.

\bibitem{6778102}
S.~Bu and F.~R. Yu, ``Green cognitive mobile networks with small cells for
  multimedia communications in the smart grid environment,'' \emph{IEEE Trans.
  on Veh. Tech.}, vol.~63, no.~5, pp. 2115--2126, June 2014.

\bibitem{7124517}
A.~Bousia, E.~Kartsakli, A.~Antonopoulos, L.~Alonso, and C.~Verikoukis,
  ``Game-theoretic infrastructure sharing in multioperator cellular networks,''
  \emph{IEEE Trans. Veh. Technol.}, vol.~65, no.~5, pp. 3326--3341, May 2016.

\bibitem{voronoi}
A.~Okabe, B.~Boots, and K.~Sugihara, \emph{Spatial Tessellations: Concepts and
  Applications of Voronoi Diagrams}.\hskip 1em plus 0.5em minus 0.4em\relax New
  York, NY, USA: John Wiley \& Sons, Inc., 1992.

\bibitem{TVTjournal1}
{H. Ghazzai, E. Yaacoub, M-S. Alouini, and A. Abu-Dayya}, ``Optimized smart
  grid energy procurement for {LTE} networks using evolutionary algorithms,''
  \emph{IEEE Trans. Veh. Technol.}, vol. 63, no. 9, pp. 4508--4519, Nov. 2014.

\bibitem{MOLPP_iterative}
J.~Matejaš and T.~Perić, ``A new iterative method for solving multiobjective
  linear programming problem,'' \emph{Elsevier Appl. Mathematics and Comput.},
  vol. 243, pp. 746--754, Sept. 2014.

\bibitem{coalitional_game_theory}
W.~Saad, Z.~Han, M.~Debbah, A.~Hjorungnes, and T.~Basar, ``Coalitional game
  theory for communication networks,'' \emph{IEEE Signal Process. Mag.},
  vol.~26, no.~5, Sep. 2009.

\bibitem{nash_equilibria}
D.~Dutta, A.~Goel, and J.~Heidemann, ``Oblivious {AQM} and {Nash} equilibria,''
  in \emph{IEEE Int. Conf. Comput. Commun. (INFOCOM 2003)}, San Francisco, CA,
  USA, Apr. 2003.

\end{thebibliography}

\begin{IEEEbiography}
    [{\includegraphics[width=1in,height=1.25in,clip,keepaspectratio]{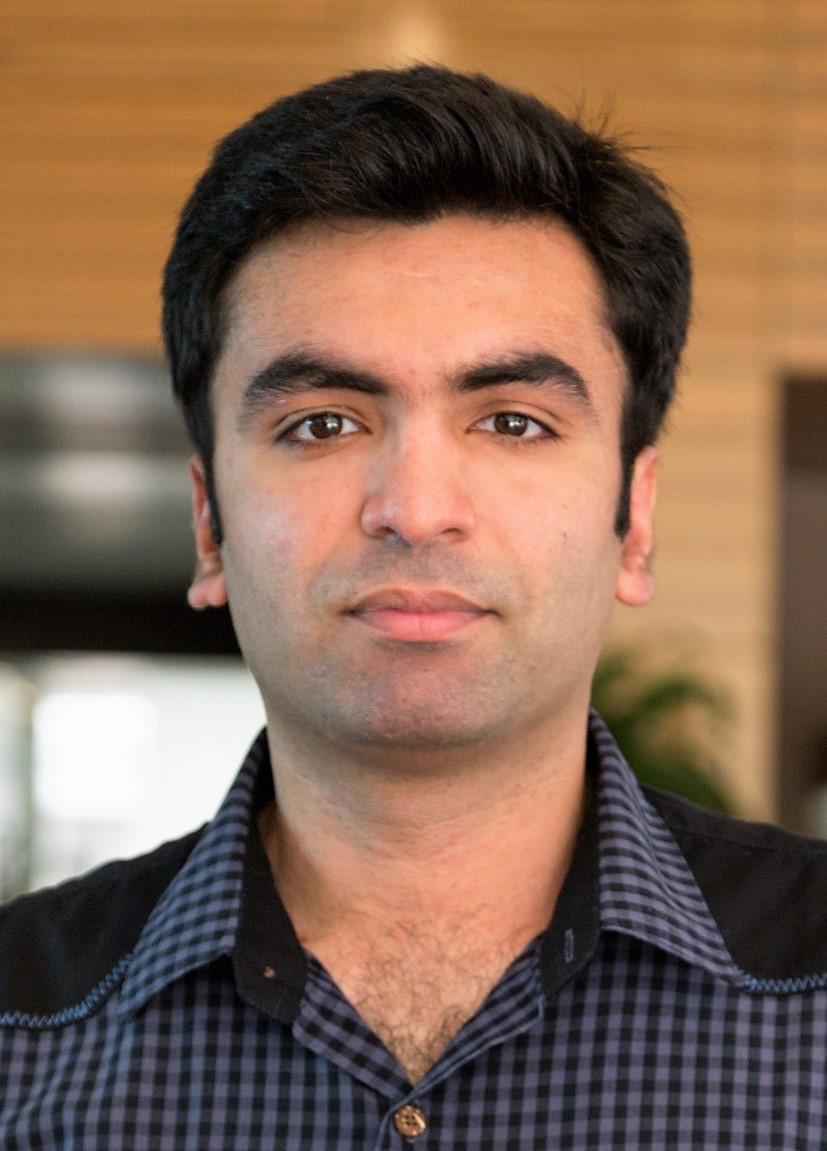}}]{Muhammed Junaid Farooq} received the B.S. degree in electrical engineering from the School of Electrical Engineering and Computer Science (SEECS), National University of Sciences and Technology (NUST), Islamabad, Pakistan, the M.S. degree in electrical engineering from the King Abdullah University of Science and Technology (KAUST), Thuwal, Saudi Arabia, in 2013 and 2015, respectively. Then, he was a Research Assistant with the Qatar Mobility Innovations Center (QMIC), Qatar Science and Technology Park (QSTP), Doha, Qatar. Currently, he is a PhD student at the Tandon School of Engineering, New York University (NYU), Brooklyn, New York. His research interests include modeling, analysis and optimization of wireless communication systems, stochastic geometry, and green communications. He was the recipient of the President's Gold Medal for the best academic performance from the National University of Sciences and Technology (NUST).
\end{IEEEbiography}

\begin{IEEEbiography}
    [{\includegraphics[width=1in,height=1.25in,clip,keepaspectratio]{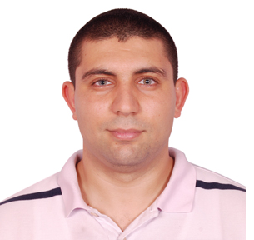}}]{Hakim Ghazzai} (S'12, M'15)
was born in Tunisia. He is currently working as a research scientist at Qatar Mobility Innovations Center (QMIC), Doha, Qatar. He received his Ph.D degree in Electrical Engineering from King Abdullah University of Science and Technology (KAUST), Saudi Arabia in 2015. He received his Diplome d'Ingenieur and Master of Science degree in telecommunication engineering from the Ecole Superieure des Communications de Tunis (SUP'COM), Tunisia in 2010 and 2011, respectively. His general research interests include mobile and wireless networks, green communications, internet of things, and UAV-based communications.
\end{IEEEbiography}

\begin{IEEEbiography}
    [{\includegraphics[width=1in,height=1.25in,clip,keepaspectratio]{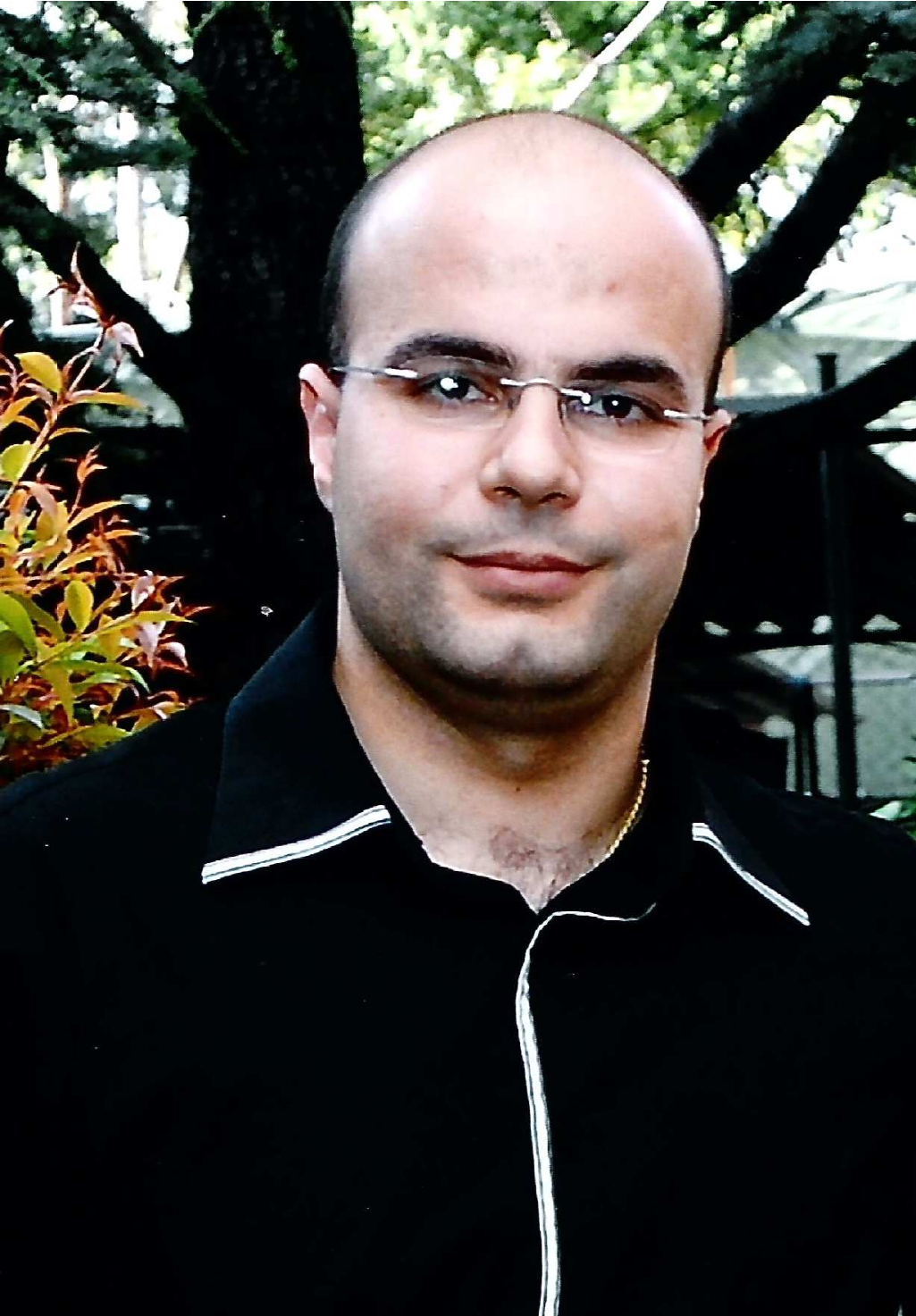}}]{Elias Yaacoub} (S'07, M'10, SM'14)
received the B.E. degree in Electrical Engineering from the Lebanese University in 2002, the M.E. degree in Computer and Communications Engineering from the American University of Beirut (AUB) in 2005, and the PhD degree in Electrical and Computer Engineering from AUB in 2010. He worked as a Research Assistant in the American University of Beirut from 2004 to 2005, and in the Munich University of Technology in Spring 2005. From 2005 to 2007, he worked as a Telecommunications Engineer with Dar Al-Handasah, Shair and Partners. From November 2010 till December 2014, he worked as a Research Scientist / R\&D Expert at the Qatar Mobility Innovations Center (QMIC). Afterwards, he joined Strategic Decisions Group (SDG) where he worked as a Consultant till February 2016. He is currently an Associate Professor at the Arab Open University (AOU). His research interests include Wireless Communications, Resource Allocation in Wireless Networks, Intercell Interference Mitigation Techniques, Antenna Theory, Sensor Networks, and Physical Layer Security.
\end{IEEEbiography}

\begin{IEEEbiography}
    [{\includegraphics[width=1in,height=1.25in,clip,keepaspectratio]{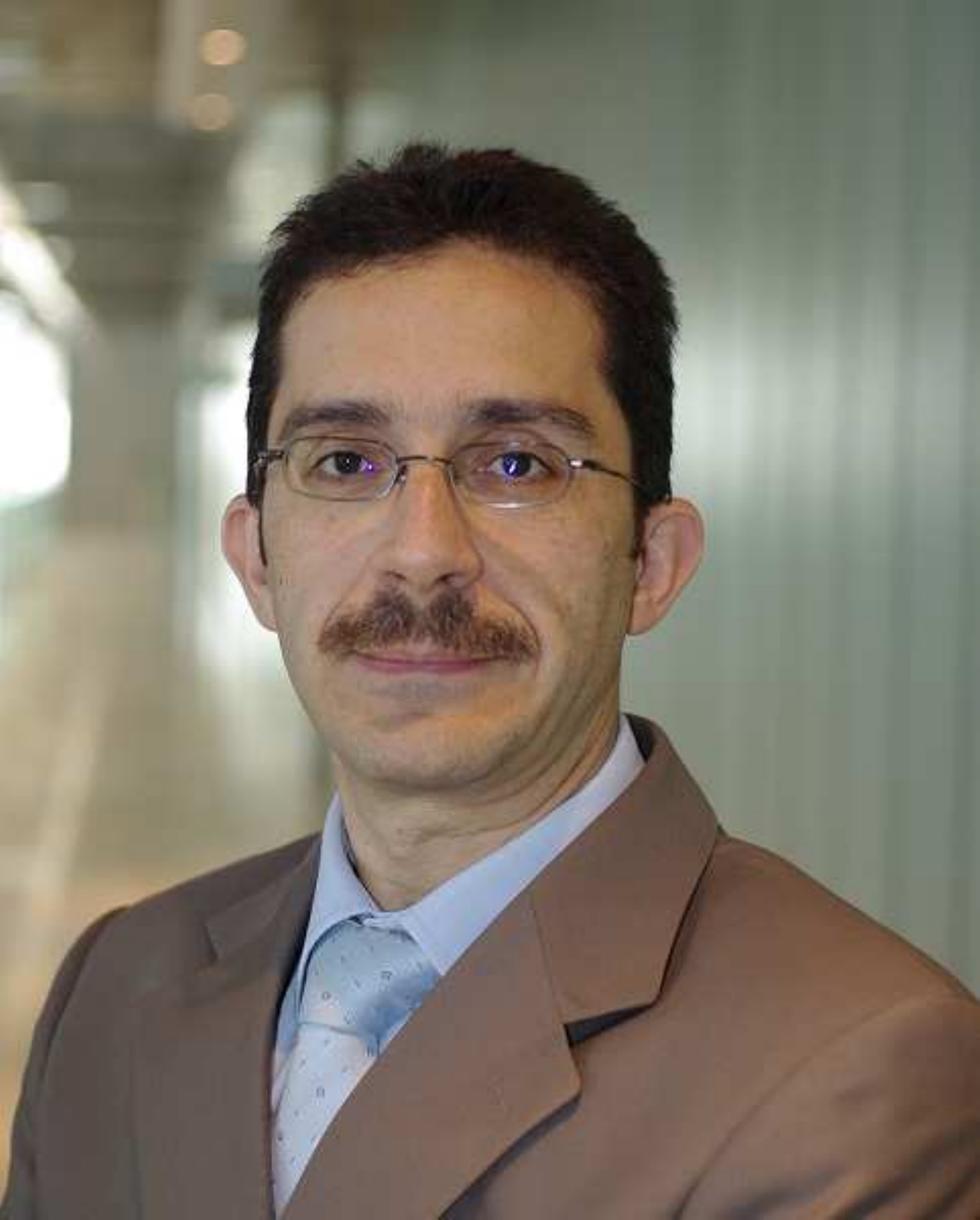}}]{Abdullah Kadri} (SM'16) received the M.E.Sc. and Ph.D. degrees in electrical engineering from the University of Western Ontario (UWO), London, ON, Canada, in 2005 and 2009, respectively. Between 2009 and 2012, he worked as a Research Scientist at Qatar Mobility Innovations Center (QMIC), Qatar University. In 2013, he became a Senior R$\&$D Expert and the Technology Lead at QMIC focusing on R$\&$D activities related to intelligent sensing and monitoring using mobility sensing. His research interests include wireless communications, wireless sensor networks for harsh environment applications, indoor localization, internet-of-things, and smart sensing. He is the recipient of the Best Paper Award at the WCNC Conference in 2014.
\end{IEEEbiography}

\begin{IEEEbiography}
[{\includegraphics[width=1in,height=1.25in,clip,keepaspectratio]{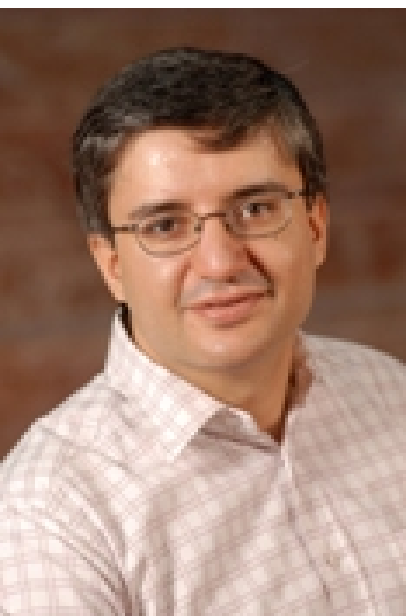}}]{Mohamed-Slim Alouini} (S'94, M'98, SM'03, F'09) was born in Tunis, Tunisia. He received the Ph.D. degree in Electrical Engineering from the California Institute of Technology (Caltech), Pasadena, CA, USA, in 1998. He served as a faculty member in the University of Minnesota, Minneapolis, MN, USA, then in the Texas A$\&$M University at Qatar, Education City, Doha, Qatar before joining King Abdullah University of Science and Technology (KAUST), Thuwal, Makkah Province, Saudi Arabia as a Professor of Electrical Engineering in 2009. His current research interests include the modeling, design, and performance analysis of wireless communication systems.
\end{IEEEbiography}

\end{document}